\documentclass[aps,prb,twocolumn,showpacs,bibliography,floatfix]{revtex4-2}
\usepackage{amsmath}
\usepackage{amsthm, amssymb} 
\usepackage{bbold}                       
\usepackage{graphicx}
\usepackage{color}
\usepackage{bm}
\usepackage{times, verbatim}      

\usepackage{ulem}

\usepackage{hyperref}
\hypersetup{       colorlinks=true, }

\newcommand{\BEq}{\begin{eqnarray}}
\newcommand{\EEq}{\end{eqnarray}}

\usepackage{academicons}
\usepackage{xcolor}
\newcommand{\orcid}[1]{\href{https://orcid.org/#1}{\textcolor[HTML]{A6CE39}{\aiOrcid}}}

\begin{document}

\title{Composite fermion mass enhancement and particle-hole symmetry of fractional quantum Hall states in the lowest Landau level under realistic conditions  }

\author{Eduardo Palacios and Michael R. Peterson\href{https://orcid.org/0000-0002-4524-6819}{\includegraphics[scale=1,width=0.225cm]{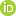}}}
\affiliation{Department of Physics \& Astronomy, California State University Long Beach,  Long Beach, California 90840, USA}

\begin{abstract}
Particle-hole symmetry breaking in the fractional quantum Hall effect has recently been studied both theoretically and experimentally with most works focusing on non-Abelian states in the second electronic Landau level. In this work, we theoretically investigate particle-hole symmetry breaking of incompressible fractional quantum Hall states in the lowest Landau level under the influence of the realistic effect of a finite magnetic field strength. A finite magnetic field induces Landau level and sub-band mixing which are known to break particle-hole symmetry at the level of the Hamiltonian. We analyze the Haldane pseudopotentials, energy spectra and energy gaps, and the wave functions themselves, under realistic conditions.  We find that particle-hole symmetry is broken, as determined by energy gaps, between states related via particle-hole conjugation, however, we find that particle-hole symmetry is largely maintained as determined by the effective mass of composite fermions. Finally, we comment and make connection to recent experimental observations regarding particle-hole symmetry in the lowest Landau level fractional quantum Hall effect [Pan \textit{et al.} Phys. Rev. Lett. 124, 156801 (2020)].
\end{abstract}

\maketitle

\section{Introduction}
The fractional quantum Hall effect (FQHE)~\cite{Tsui82} has continued to fascinate and serve as a platform to explore exotic two-dimensional topological phases of matter under extreme quantum conditions since its experimental discovery in 1982~\cite{Prange90,DasSarma97,jain2007composite}.  A particular driver for recent research, both theoretical and experimental, is the possibility of topologically ordered phases supporting non-Abelian excitations with applications in fault-tolerant quantum information processing~\cite{Kitaev97,Freedman01,Nayak08}.  
 
Research on the issue of particle-hole invariance of FQHE states has focused mostly on the FQHE in the half-filled second Landau level~\cite{Willett87}(i.e., filling factor $\nu=\frac{5}{2}$) but additional work has studied other exotic non-Abelian states such as the one at $\nu=\frac{12}{5}$~\cite{Xia04}.  In the limit of infinite magnetic field strength, the Hamiltonian for the FQHE is invariant under particle-hole conjugation but finite magnetic field induced Landau level mixing breaks this symmetry explicitly~\cite{Bishara09a,Peterson13b,Sodemann13,Simon13,Wooten13}.  It is, therefore, a matter of details and careful calculation to determine whether a realistic system will host phenomena that is invariant under particle-hole conjugation--we refer the reader to  work done recently in the second Landau level~\cite{Wojs10,Rezayi11,Son15,PetersonPRB2015,Pakrouski15,Zaletel15,Pakrouski16,Rezayi17,Schreiber18,Banerjee18,Mross18,Wang18,Lian18,Simon18,Ma19,Zhu19,Simon20}.

In this work, we study particle-hole conjugate  FQH states in the lowest Landau level.  Our primary motivation is twofold.  First, it is clear that particle-hole symmetry is broken explicitly via Landau level mixing due to finite magnetic field strength at the level of the Hamiltonian and that this symmetry breaking has consequences for FQHE states in the second Landau level, i.e., it  apparently energetically favors the anti-Pfaffian~\cite{Levin07,Lee07,Rezayi17} over the Moore-Read Pfaffian~\cite{Moore91}, and renders the potential FQHE state at $\nu=\frac{13}{5}$ energetically unfavorable compared with  the FQHE state at $\nu=\frac{12}{5}$~\cite{Zhu15,Pakrouski16,Mong17}--in fact, the $\nu=\frac{12}{5}$ FQHE has been experimentally observed~\cite{Xia04,Kumar10,Choi08,Pan08,Zhang12,Deng12} while the $\nu=\frac{13}{5}$ has not.  One motivation for the current work is to study theoretically FQHE states under particle-hole conjugation in the lowest Landau level in a system for which the Hamiltonian has explicitly broken particle-hole symmetry due to realistic effects.  There is a relative lack of prior numerical work studying particle-hole symmetry in the lowest LL with most recent work focusing on half-filling~\cite{Balram17,Sreejith2017}. Reference~\onlinecite{Sreejith2017}, however, did study the filling factors $\nu=\frac{1}{3}$ and $\frac{2}{3}$ using fixed phase diffusion Monte Carlo and found that particle-hole symmetry between the two, measured via energy gaps, was maintained for strong LL mixing.   Our results presented below using exact diagonalization do not agree with the fixed phase diffusion Monte Carlo results and  we do not speculate on possible reasons.  As mentioned briefly above, Ref.~\onlinecite{Pakrouski16} studied LL mixing effects between $\nu=\frac{12}{5}$ and $\frac{13}{5}$ in the second LL and compared the physics with that of the lowest LL by calculating the relative energy-gap difference and wave-function overlap between particle-conjugate states at $\nu=\frac{2}{5}$ and $\frac{3}{5}$ for very weak LL mixing finding the two to be nearly the same.  Finally, Ref.~\onlinecite{PetersonPRB2015} studied the $\nu=\frac{2}{3}$ bilayer problem under realistic conditions and incidentally found that the wave function overlap between the exact ground state in the single-layer limit (strong tunneling) and the composite fermion state was nearly unity under LL mixing.

The second motivation is experimental.  While there is a plethora of experimental observations of FQHE states in the lowest Landau level going back nearly 30 years, a recent work by Pan \textit{et al}.~\cite{Pan20} studied it explicitly.  One reason for this historical lack of specific experimental studies on particle-hole invariance in the lowest Landau level is that the measurement and interpretation of, for example, energy gaps at $\nu$ and its particle-hole conjugate filling factor $1-\nu$ is complicated by the fact that typically the two states are measured at different magnetic-field strengths at fixed density.  However, Pan, \textit{et al}.~\cite{Pan20} systematically measured FQHE gaps at $\nu$ and $1-\nu$ at a wide range of fixed magnetic-field strengths  from $B=4.5-13.5$ T, ultimately finding that particle-hole invariance between FQHE states at $\nu$ and $1-\nu$ is evidently obtained, as measured by energy gaps, for  all FQHE states studied in the lowest Landau level.  

Using exact diagonalization in the spherical geometry we study an effective Hamiltonian that includes LL mixing effects perturbatively, to lowest order, in the LL mixing parameter parameter $\kappa=(e^2/\epsilon\ell_0)/\hbar\omega_c$ which is the ratio of the Coulomb energy ($e^2/\epsilon\ell_0)$ to the cyclotron energy ($\hbar\omega_c$) and depends inversely on the square root of magnetic-field strength $B$.  As mentioned above, this effective Hamiltonian has been studied in the context of the second LL but, so far, has not been systematically examined in the lowest Landau level.  We find that particle-hole invariance is broken, but in a subtle way: the energy gaps behave quantitatively differently under LL mixing for FQHE states related via particle-hole conjugation with the smaller filling factor energy gaps decreasing more under increasing LL mixing. Meanwhile, the ground states at $\nu$ and $1-\nu$ are extremely similar as measured by wave-function overlap while the first-excited state shows some differences.  Finally, we find the effective masses for composite fermions at fillings $\nu$ and $1-\nu$ are enhanced by realistic effects, are in reasonable agreement with recent measurements by Pan, \textit{et al}.~\cite{Pan20}, and are approximately particle-hole invariant.

The plan of this work as as follows.  In Sec.~\ref{sec:ham} we introduce the spherical geometry, the effective Hamiltonian, and 
discuss the finite-size systems we study and limitations therein.  Section~\ref{sec:gaps} studies FQHE energy gaps, wave-function overlaps between exact ground states and composite fermion states, an examination of the two-body and three-body pseudopotentials, and a quantitative measure of particle-hole symmetry breaking.  Finally in sections~\ref{sec:exp} and~\ref{sec:conc}, we compare our theoretical results with experimental results and conclude.

\section{Effective Hamiltonian}
\label{sec:ham}
The Hamiltonian for the FQHE is
\begin{equation}
\hat{H} = \hat{H}_{0} + \frac{e^2}{\epsilon}\sum_{i<j}\frac{1}{|\bf{r}_j-\bf{r}_k|}+\sum\limits_{j}\hat{U(\bf{r}_j)}\;,
\end{equation}
where $\hat{H_0}$ is the kinetic-energy term defining the Landau levels, the final term consists of single-particle terms such as confining potentials, etc., and $e$ is the electron charge and $\epsilon$ is the dielectric of the host semiconductor, which in typical GaAs/AlGaAs heterostructures is $\epsilon\approx 12.6$.  In this work the system is assumed to be spin-polarized, so Zeeman effects are neglected. The middle term is the Coulomb interaction, which for the FQHE, dominates all the action.

Following Refs.~\onlinecite{Bishara09a,Peterson13b,Sodemann13},  an effective Hamiltonian that includes LL mixing perturbitavely and projected onto to the $n$-th Landau level  is written as
\begin{align}
    \hat{H}_{eff}(\kappa) = \sum\limits_{m}[V_{m}^{(2)}(n)+\kappa\delta V_{m}^{(2)}(n)]\sum\limits_{i<j}\hat{P}_{ij}(m)\nonumber \\
    +\sum\limits_{m}\kappa V_{m}^{(3)}(n)\sum\limits_{i<j<k}\hat{P}_{ijk}(m)\;,
 \label{Hammy}
\end{align}
where the $V_{m}^{(2)}(n)$ is the bare two-body Haldane pseudopotential~\cite{Haldane83} in the $n$-th LL and $\delta V_{m}^{(2)}(n)$ is the correction term due to Landau-level mixing, $V_{m}^{(3)}(n)$ are emergent three-body pseudopotentials, and $\hat{P}_{ij}(m)$ and $\hat{P}_{ijk}(m)$ are two and three-body projection operators, respectively. Landau-level mixing is parametrized by $\kappa$ which, in the case of GaAs heterostructures, $\kappa= (e^2/\epsilon \ell_0)/\hbar \omega_c\approx 2.52/\sqrt{B}$  where $\omega_c=eB/mc$ is the cyclotron frequency ($m\approx 0.067m_e$ is the band mass of the electrons), and $\ell_0 =\sqrt{\hbar c / e B} $  is the magnetic length. Thus, as $\kappa\rightarrow 0$ (or $B\rightarrow\infty$) the effective Hamiltonian is exact.  With the Hamiltonian constructed, we investigate the gaps, wave-function overlaps, and particle-hole invariance via exact diagonalization within the spherical geometry. 

The composite fermion theory of the FQHE ~\cite{Jain89,jain2007composite} well describes the over 70 experimentally observed FQH states in the lowest Landau level at filling factor $\nu=n/(2pn\pm 1)$.  Therefore, we appeal to it to guide the present study.  Briefly, the composite fermion theory posits that strongly interacting electrons with band mass $m$ and density $\rho$ in the presence of an applied magnetic-field strength $B$ transform into weakly interacting composite fermions with effective mass $m^*$ in order to minimize the Coulomb interaction energy between electrons~\cite{Jain89,jain2007composite}.  A composite fermion is thought of as an electron bound to an even number ($2p$, $p>0$ integer) of vortices of the many-body wave function--these are often heuristically imagined as flux quanta of the applied magnetic field.  This binding of vortices amounts to the composite fermions experiencing a reduced effective magnetic field of strength $B^*=B - 2p\rho\phi_0$ ($\phi_0 = hc/e$ is the magnetic flux quanta).  When the density is such that composite fermions  fill an integer number $\nu^*=n$ of composite fermion LLs, called $\Lambda$-levels in the literature, the system has an energy gap and the phenomena of the quantum Hall effect is obtained.  The composite fermion filling factor $\nu^*=\rho/(B^*/\phi_0)$ is related to the electron filling factor via $\nu=\nu^*/(2p\nu^*+1)$.  Thus, the FQHE of electrons is understood as the integer quantum Hall effect of composite fermions.

In our calculations we use the spherical geometry since it is a compact geometry without edges and ideal to study energy gaps for FQHE states at filling factor $\nu=n/(2pn\pm 1)$.  Thus, the two-dimensional electron system in the presence of a perpendicular constant magnetic field $B$ is mapped to a system where a magnetic monopole of strength $Q$ is placed at the center of a sphere of radius $R=\sqrt{Q}\ell_0$ with $N$ electrons on the surface.  The monopole produces a radial magnetic field of constant strength with total magnetic flux through the surface of $2Q\phi_0$.  We use the composite fermion theory to fix the particle number $N$ and monopole strength $Q$ in our exact diagonalization studies.

All FQH states for filling factor $\nu$ have  a relationship between the total flux $2Q$ and particle number $N$ as $2Q = \nu^{-1}N - \chi$, where $\chi$ is the so-called  topological  shift related to the specific topological order of the FQH state~\cite{Wen90a}.  Note that, for simplicity and due to numerical constraints, we fix $p=1$  and only consider composite fermions consisting of electrons bound to two vortices of the many-body wave function. Composite fermion theory gives $2Q = 2Q^* + 2(N-1)$ where $2Q^* = N/n - n$, i.e., 
\BEq
2Q &=&\nu^{-1}N - (n+2)\,,
\EEq
 and, hence defines the  shift as $\chi = n + 2$. In this work, we are also interested in the particle-hole conjugate  of the FQHE states.   Under PH conjugation $N\rightarrow N_h$, where $N_h$ is the number of holes, yielding
\BEq
2Q = (1-\nu)^{-1}N_h - \left(\frac{1-\nu\chi}{1-\nu}\right)
\EEq
where we used $2Q+1=N+N_h$ with $2Q+1$  equaling the finite LL degeneracy in the spherical geometry.

Table \ref{tab:dim_table} shows the various systems studied in this work along with the dimension of the Hilbert space of each. The Hilbert spaces grow exponentially, making exact diagonalization difficult to execute  for increasingly larger system sizes. Due to this limitation only systems with dimensions under a million were considered for the filling fractions shown -- this limitation is largely due to the fact that the three-body projection operators are more dense than the two-body projection operators.

\begin{table}[!h]
\caption{\label{tab:dim_table} Systems studied along with their various filling fractions and Hilbert-space dimension (dim $\mathcal{H}$).  Larger systems were not studied due to the prohibitive size of the Hilbert space or due to aliasing with other FQH states which complicates the interpretation of the results.}
\begin{ruledtabular}
\begin{tabular}{lcr}
 & $\nu(1-\nu) = \frac{1}{3}(\frac{2}{3})$ & \\ 
\hline
\hline
$N(N_h)$ & $Q$ & $\dim{\mathcal{H}}$  \\
\hline
4(6) & 4.5 & 18      \\
6(10) & 7.5 & 338      \\
7(9) & 9 & 1,656        \\
8(14) & 10.5 & 8,512      \\
9(16) & 12 & 45,297       \\
10(18) & 13.5 & 246,448  \\
\hline\hline
 & $\nu(1-\nu) = \frac{2}{5}(\frac{3}{5})$ & \\ 
\hline
\hline
$N(N_h)$ & $Q$ & $\dim{\mathcal{H}}$  \\
\hline
 4(3) & 3 & 5       \\
 6(6) & 5.5 & 58  \\
8(9) & 8  & 910    \\
 10(12) & 10.5 & 16,660     \\
 12(15) & 13 & 332,578  \\
\end{tabular}
\end{ruledtabular}
\end{table}

\section{Gaps, Overlaps, Pseudopotentials and Particle-hole Symmetry Breaking}
\label{sec:gaps}

\subsection{Pseudopotential Analysis}

\begin{figure}[!h]
    \includegraphics[width=3.4in]{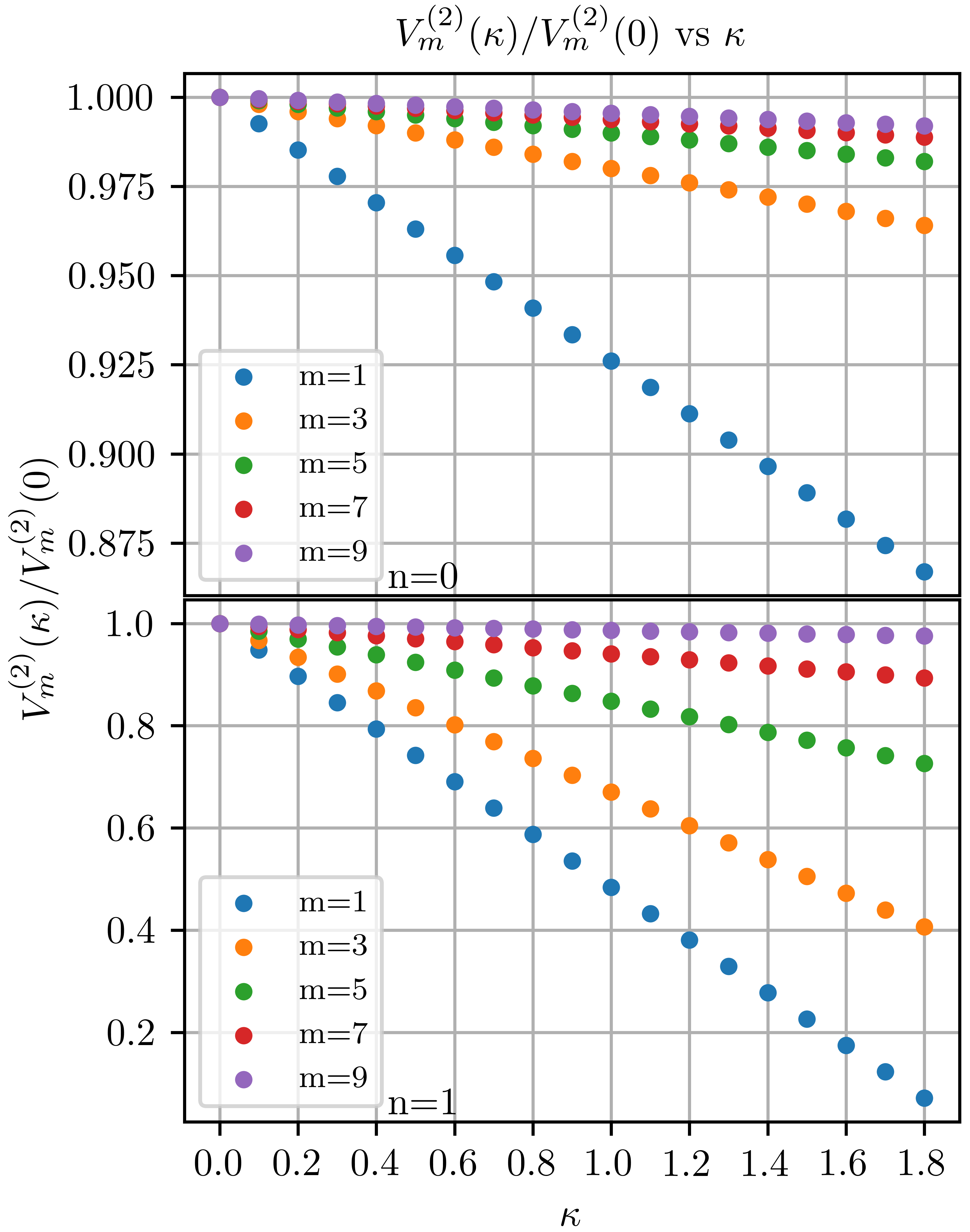}
\caption{Two-body pseudopotential ratios, $V_m^{(2)}(\kappa)/V_m^{(2)}(0)$ (we have suppressed the LL index $n$ in the argument), for the lowest Landau level ($n=0$) (top), and the second Landau Level ($n=1$) (bottom). The lowest LL pseudopotential values change much less than the second LL ones.  Note that as $m$ increases the effect of LL mixing is diminished since larger $m$ corresponds to larger spatial separation between electrons.}
\label{fig:V2}
\end{figure}

To obtain some qualitative understanding of the physics of LL mixing we begin by examining the  values for both the $V^{(2)}_{m}(n,\kappa)$ and $V^{(3)}_{m}(n)$ Haldane pseudopotentials where $V^{(2)}_{m}(n,\kappa)\equiv V_{m}^{(2)}(n)+\kappa\delta V_{m}^{(2)}(n)$ is the full two-body pseudopotential under LL mixing. Figure \ref{fig:V2} shows the ratio of the $\kappa$ dependent two-body pseudopotentials to the bare two-body pseudopotentials for the first five odd angular-momentum quantum numbers in the lowest two LLs (only odd $m$ pseudopotentials are important for spin-polarized systems). Obviously, as $\kappa$ grows the ratios decrease to zero linearly in $\kappa$.  However, since $\delta V^{(2)}(1)$ have larger absolute values compared with $\delta V^{(2)}(0)$, cf. Ref.~\onlinecite{Peterson13b}, the LL mixing pseudopotentials for the lowest LL are much less sensitive to the changes in $\kappa$ than those for the second LL. For example, when $\kappa=1$, or when $B\approx$ 6.4 T for GaAs systems, the $m=1$ LLL pseudopotential is approximately $~8\%$ smaller than its $\kappa=0$ value, whereas in the second LL it is $~50\%$ smaller. This qualitative difference between the pseudopotentials in the two LLs is independent of $m$ as shown in Figure \ref{fig:V2} (note the scale of the ordinate axis).    Thus, from the two-body pseudopotentials alone, one might expect LL mixing to have a weaker effect in the LLL compared with the second LL. 

In Table~\ref{table:V3} we investigate the three-body pseudopotentials $V^{(3)}_{m}$ that emerge from including LL mixing in the LLL and compare them to those in the second LL. Importantly, the three-body terms are those that break particle-hole symmetry. For $m\in$[3,6,8], the LLL pseudopotentials are negative which result in attractive forces that are directly proportional to $\kappa$. The rest of the included pseudopotentials are repulsive and become even more so as $\kappa$ is increased. Investigating the physics of particle-hole symmetry-breaking three-body terms is complicated by our lack of intuition regarding these terms.  However, general comments can be made.  First, the absolute value of $V_m^{(3)}(n)$ are similar for $n=0$ and $n=1$, as shown in the ratio of the two given in the last column in Table~\ref{table:V3}.  We note that when the ratio is not near one, the pseudopotentials themselves are numerically very small.  Second, we note that the three-body pseudopotentials are approximately an order of magnitude smaller than the full two-body pseudopotentials [$V^{(2)}_m(n,\kappa)$] yet are the driver for particle-hole symmetry breaking.

\begin{table}[!h]
\caption{\label{table:V3} Three-body pseudopotential values in the lowest and second LL.}
\begin{ruledtabular}
\begin{tabular}{cccc}
$m$ & $V^{(3)}_{m}(n=0)$ & $V^{(3)}_{m}(n=1)$ & $\frac{V^{(3)}_{m}(n=1)}{V^{(3)}_{m}(n=0)}$ \\
\hline
3          & -0.018101                 & -0.014684                 & 0.811226     \\ 
5          & 0.003265                  & -0.005352                 & -1.63920     \\ 
6          & -0.010681                 & -0.009899                 & 0.926786     \\ 
7          & 0.005945                  & 0.000451                  & 0.075862    \\ 
8          & -0.004714                 & -0.000945                 & 0.200467                                          
\end{tabular}
\end{ruledtabular}
\end{table}

\subsection{Energy gaps}

\begin{figure}[!h]
\includegraphics[width=3.4in]{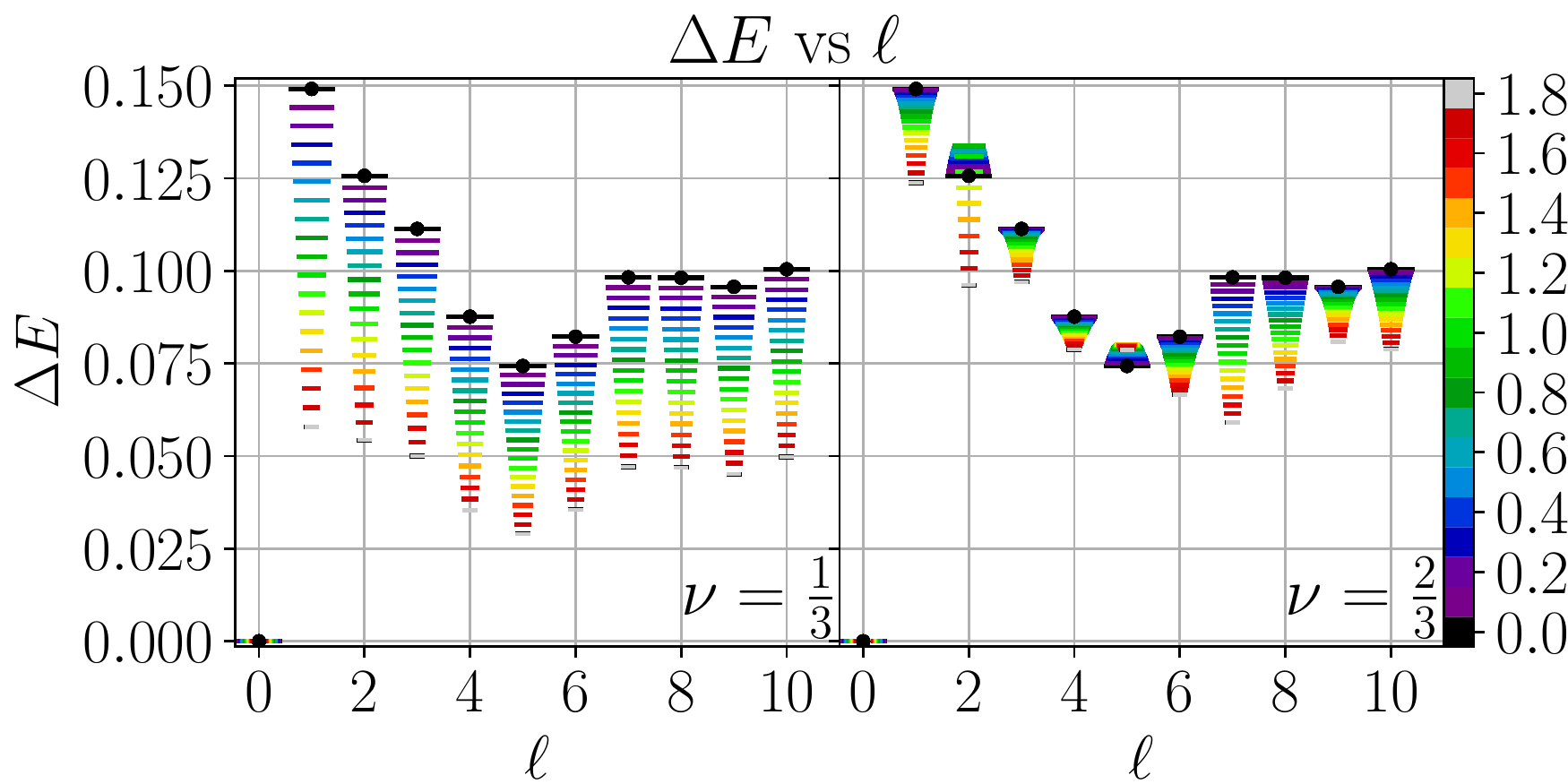}\\
\includegraphics[width=3.4 in]{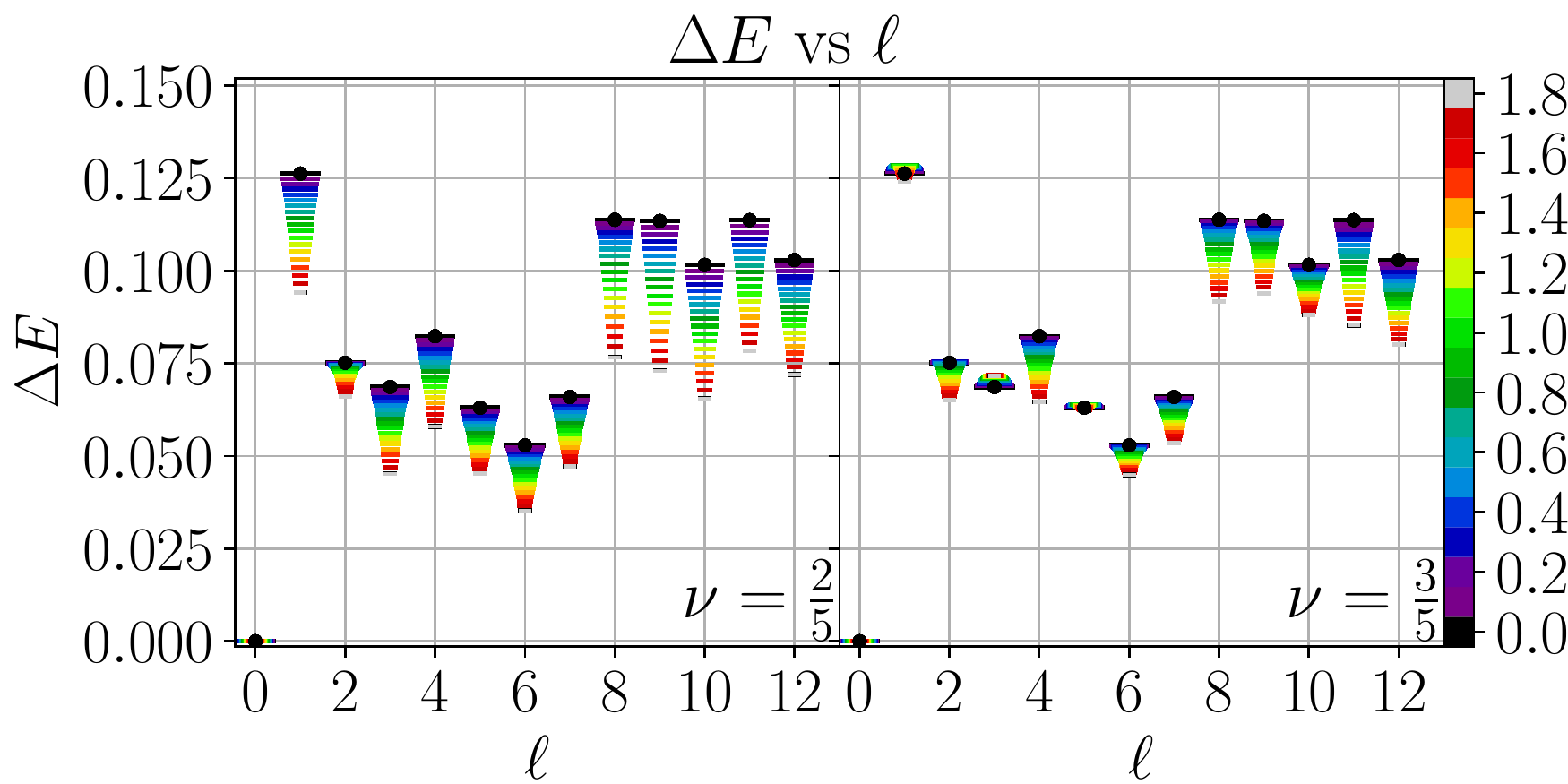}
\caption{Energy spectra for $N=10(18)$ and $Q=13.5$ corresponding to $\nu=\frac{1}{3}$ and $\frac{2}{3}$ (top), respectively, and for $N=12(15)$ and $Q=13.0$ corresponding to $\nu=\frac{2}{5}$ and $\frac{3}{5}$ (bottom), respectively, as $\kappa$ is varied. The solid black circles represent the cases when $\kappa =0$ which produces identical spectra for both graphs.  As $\kappa$ is increased, particle-hole symmetry breaking is manifest--the amount of particle-hole symmetry breaking is less pronounced for $\nu=\frac{2}{5} (\frac{3}{5})$ compared to $\nu=\frac{1}{3} (\frac{2}{3})$.}
\label{fig:thirds}
\end{figure}

The previous section investigated the physics of LL mixing in the lowest LL at the single-particle level by investigating the pseudopotentials.  We now discuss our results from the exact diagonalization in terms of energy spectra and gaps, which are directly observable experimentally.  In the following figures we plot the lowest branch of excited states with respect to the ground state as a function of total angular momentum $l$.  Excluding Landau-level mixing yields a Hamiltonian constructed only of two-body operators, which are naturally particle-hole symmetric. This symmetry is manifest in Fig.~\ref{fig:thirds} for the filling factors $\nu=\frac{1}{3}$ and $1-\nu=\frac{2}{3}$, and  $\nu=\frac{2}{5}$ and $1-\nu=\frac{3}{5}$; the solid black circles on the energy spectra indicate the case when $\kappa = 0$ which excludes Landau-level mixing. Figure \ref{fig:thirds} also shows energy spectra of the lowest branch of excitations for $\kappa\in[0,1.8]$ where each value of $\kappa$ is represented by a color; ``warmer" colors correspond to larger values of $\kappa$. In addition, the length of each  line used for each energy also decreases with increasing $\kappa$ so that the collection of spectra across $\kappa$ depicts the general direction of change for each $l$  by drawing an ``arrow".

\begin{figure}
\includegraphics[width=3.4in]{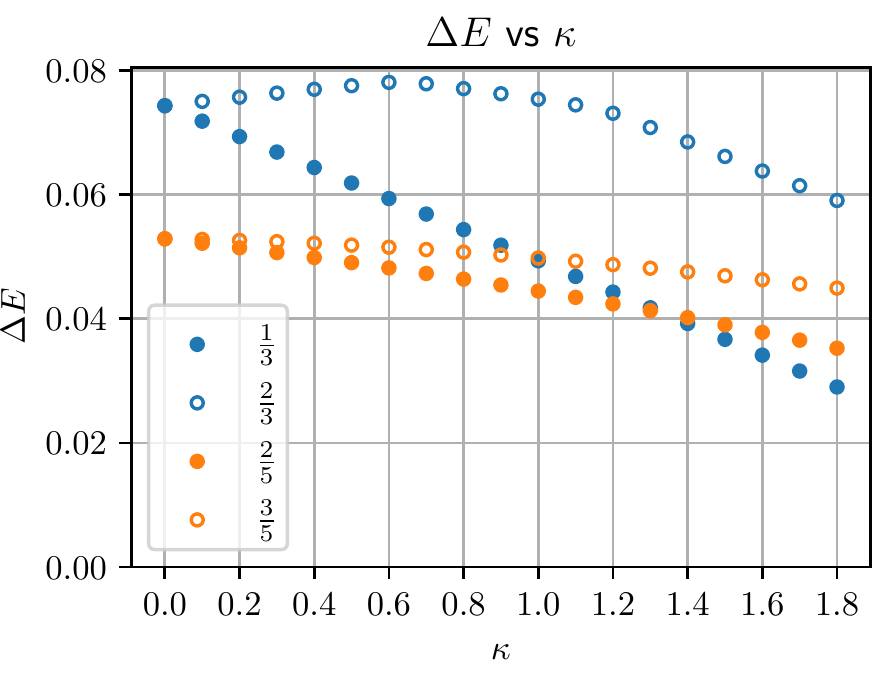}\\
\includegraphics[width=3.4in]{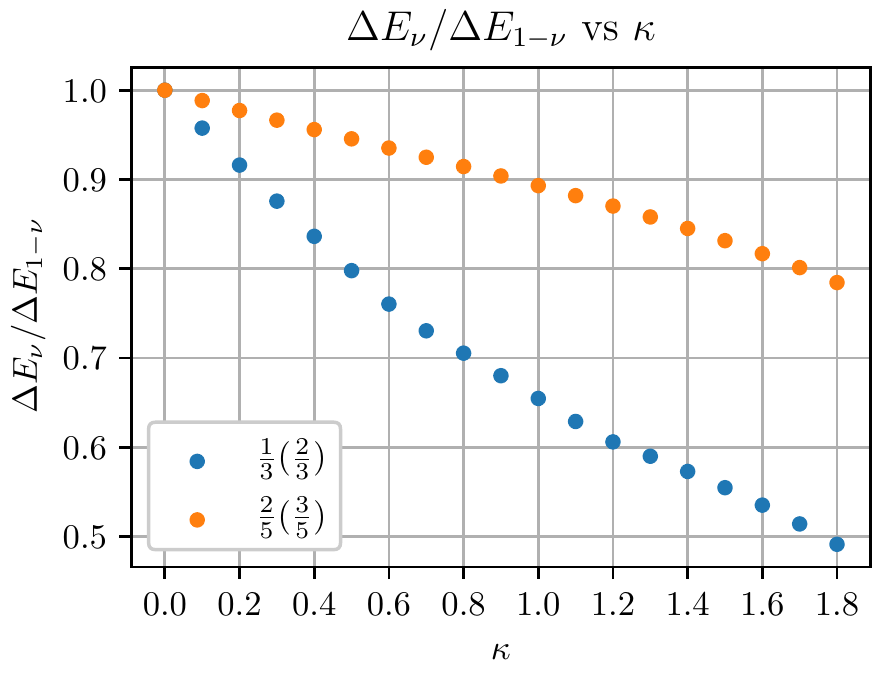}
\caption{(top) Energy difference $\Delta E$ between the first-excited state and the ground state as a function of $\kappa$ for $N=10(18)$ and $Q=13.5$ for $\frac{1}{3}(\frac{2}{3})$ and $N=12(15)$ and $Q=13.0$ for $\frac{2}{5}(\frac{3}{5})$. The gaps decrease as $\kappa$ increases, and the gaps for $\nu$ and $1-\nu$ diverge from each other indicating that the particle-hole symmetry, as measured by the energy gap, is broken under LL mixing. (bottom) The ratio $\Delta E_\nu/\Delta E_{1-\nu}$ for $N=10(18)$ and $Q=13.5$ for $\frac{1}{3}(\frac{2}{3})$ and $N=12(15)$ and $Q=13.0$ for $\frac{2}{5}(\frac{3}{5})$, respectively, shows that $\frac{1}{3}$ and $\frac{2}{5}$ both have smaller gaps than their particle-hole conjugates for any $\kappa$. }
\label{fig:gaps}
\end{figure}

As $\kappa$ is increased,  the energy spectra for $\nu$ and $1-\nu$ diverge from each other. The smaller filling fractions ($\nu$) drop in a monotonic manner as $\kappa$ increases compared with the larger filling factors $1-\nu$ which exhibit more complicated behavior. Figure \ref{fig:gaps} shows the energy difference $\Delta E$ between the ground state and the first-excited state, i.e., the so-called neutral gap, which is typically the roton minimum, cf. Ref.~\onlinecite{jain2007composite}. As $\kappa$ is increased, the gaps begin to decrease approximately linearly for $\frac{1}{3}$ and $\frac{2}{5}$, whereas  both the $\frac{2}{3}$ and $\frac{3}{5}$ cases are more resistant to LL mixing effects. The lower panel in Fig.~\ref{fig:gaps} shows the ratios of the energy gaps at $\nu$ to $1-\nu$ indicating the particle-hole symmetry is broken more severely between $\nu=\frac{1}{3}$ and $\frac{2}{3}$ than it is between $\frac{2}{5}$ and $\frac{3}{5}$. Evidently, $\Delta E_\nu < \Delta E_{1-\nu}$ and the particle-hole symmetry is broken less as $\nu$ and $1-\nu$ increase or decrease, respectively.

\begin{figure}[htp]
\includegraphics[width=3.4in]{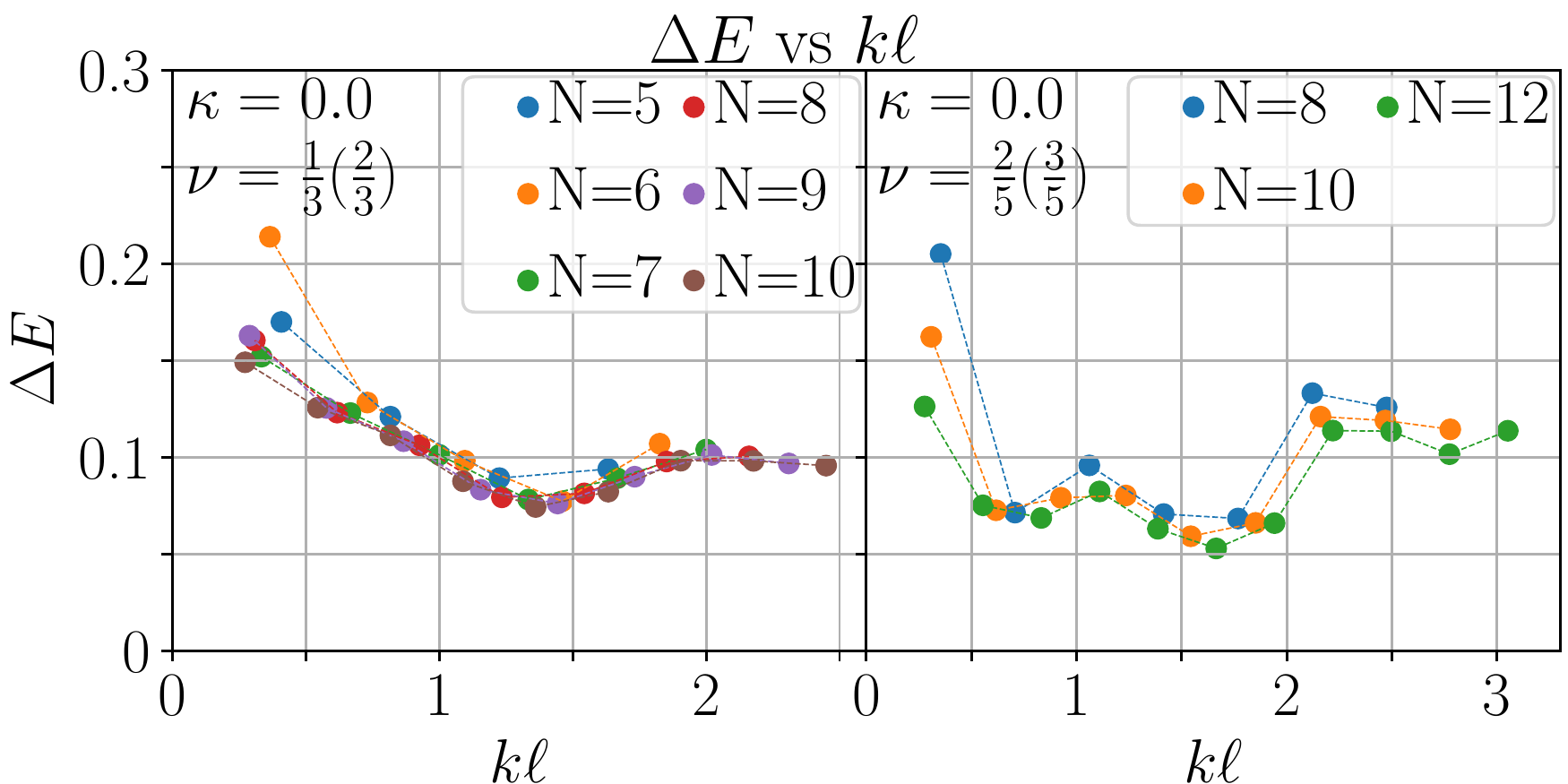}
\caption{Low energy dispersion at $\kappa$=0 for $\nu=\frac{1}{3}(\frac{2}{3})$ (left panel) and $\nu=\frac{2}{5}(\frac{3}{5})$ (right panel) as a function of wave vector $kl=l/\sqrt{Q}$.}
\label{fig:therm_unpertub}
\end{figure}

To connect the theoretically calculated energy gaps to those observed in experiment it is important to examine the thermodynamic limit.  A good indicator of whether the thermodynamic limit has been approximated is to check the energy dispersion, i.e., energy versus momentum, as we increase system size. To this end we convert the total angular momentum $l$ in Fig.~\ref{fig:thirds}  to wave vector $kl=l/\sqrt{Q}$. Figure \ref{fig:therm_unpertub} shows the lowest branch of the energy dispersion for various system sizes when $\kappa=0$. Since this case is naturally particle-hole symmetric, both $\nu$ and $1-\nu$ share the same energy dispersion. As the system size grows, the lowest energy branches approximately converge onto a single curve indicating that the largest systems sizes provide an adequate approximation of the energy gap in the thermodynamic limit.

\begin{figure}[h!]
\includegraphics[width=3.4in]{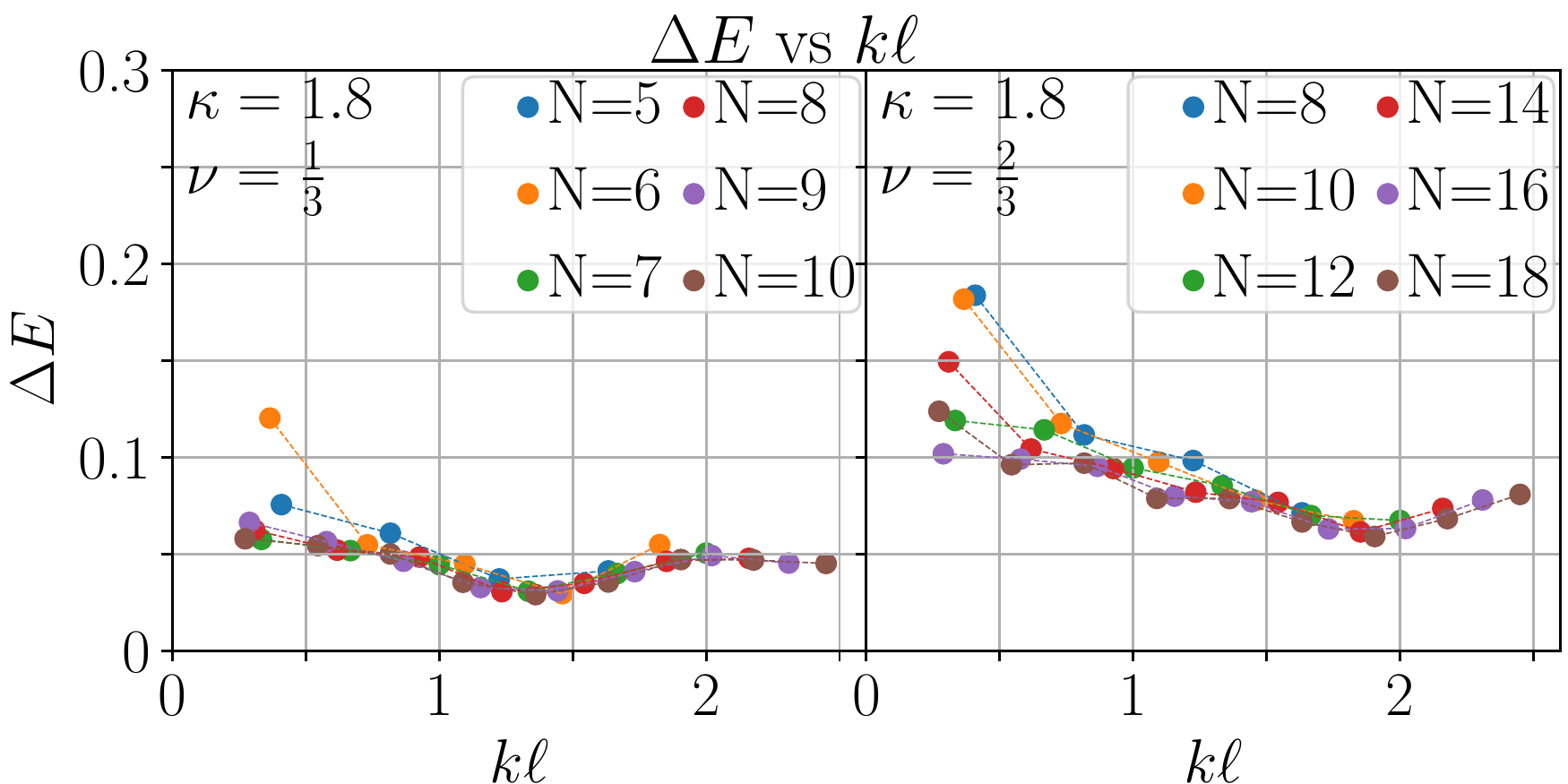}\\
\includegraphics[width=3.4in]{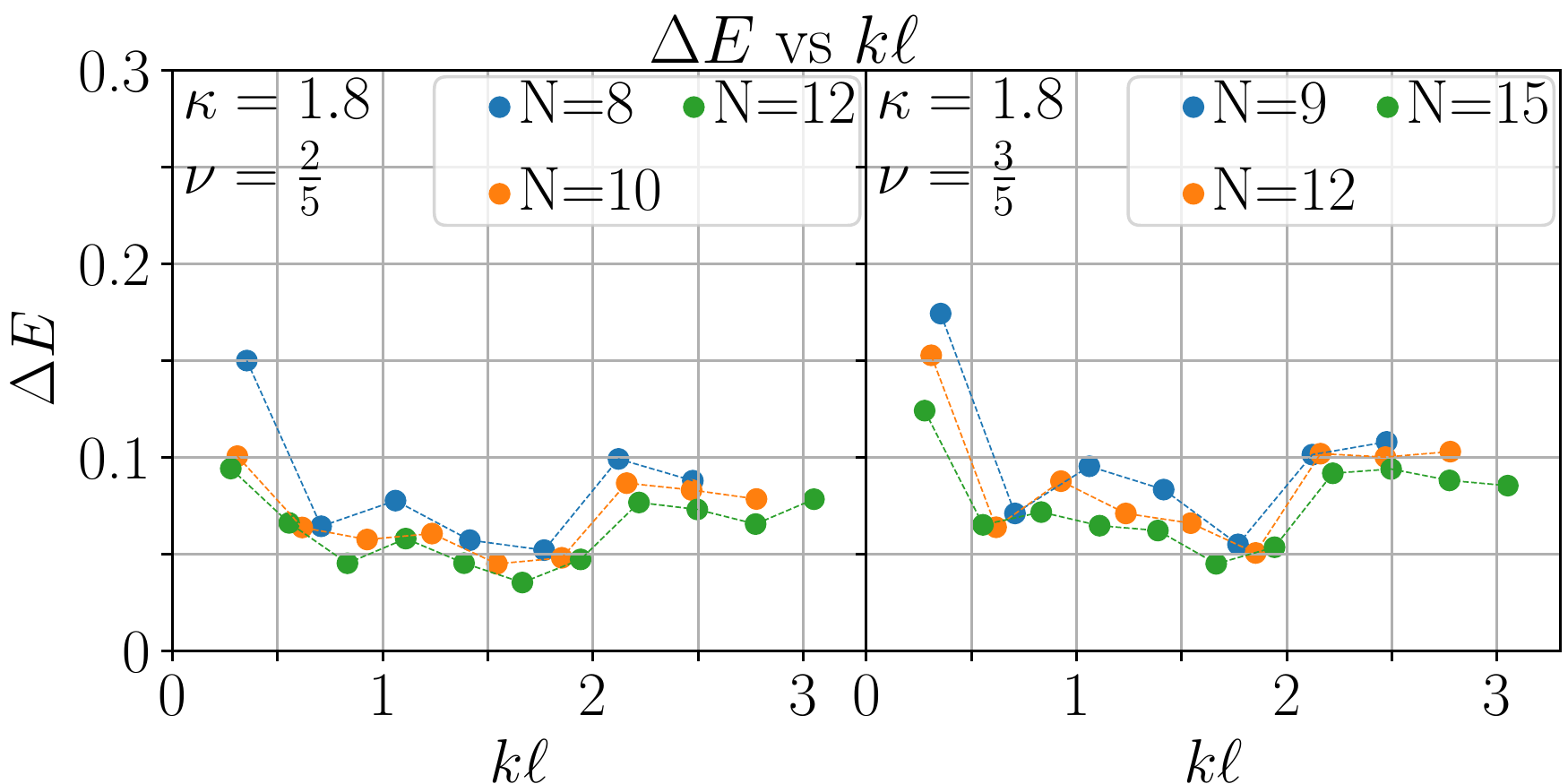}
\caption{Low energy dispersion for $\nu=\frac{1}{3}(\frac{2}{3})$ (top) and $\nu=\frac{2}{5}(\frac{3}{5})$ (bottom) for $\kappa=1.8$.}
\label{fig:therm_thirds}
\end{figure}

The thermodynamic limit is approximately maintained also for finite $\kappa$. Figure~\ref{fig:therm_thirds} shows the low energy dispersion  for $\nu=\frac{1}{3} (\frac{2}{3})$ and $\nu=\frac{2}{5} (\frac{3}{5})$ at $\kappa=1.8$, respectively. We can see that even at $\kappa > 1$ the systems remain close to the thermodynamic limit for both cases. However, the thermodynamic limits are different than they are for $\kappa=0$ which is evidence of particle-hole symmetry breaking of the energy gap even in the thermodynamic limit. We note again, that while there is evident particle-hole symmetry breaking between $\nu=\frac{2}{5}$ and $\frac{3}{5}$ it is much less severe compared with $\nu=\frac{1}{3}$ and $\frac{2}{3}$.

From an analysis of the energy spectra and energy gaps we see that particle-hole symmetry is broken under LL mixing in the lowest LL.  However, the particle-hole symmetry is broken more severely for small filling factors, i.e., it is broken more for $\nu=\frac{1}{3} (\frac{2}{3})$ than it is for $\nu=\frac{2}{5} (\frac{3}{5})$.  In addition, while LL mixing generically reduces all energy gaps for FQH states, it reduces the gap at $\nu$ more compared with the gap at $1-\nu$.  To understand more how the particle-hole symmetry is broken we now analyze wave-function overlaps.

\subsection{Overlaps}

In this section we analyze the particle-hole symmetry of the ground-state and first-excited state wave functions themselves in order to better understand the apparent particle-hole symmetry breaking observed in the FQHE energy gaps.  More precisely we compute the overlap between the ground state $|\Psi^\nu_0\rangle$ at $\nu$ (the state with $N$ electrons and $2Q$) and the particle-hole conjugate of the state $\mathrm{conj}(|\Psi^{1-\nu}_0)\rangle$ at $1-\nu$ (the state with $2Q+1-N$ electrons at flux $2Q$)--this type of measure was previously used to investigate particle-hole symmetry breaking in Ref.~\onlinecite{Pakrouski16}, for example.  If the states at $\nu$ and $1-\nu$ are related simply by particle-hole conjugation, then the overlap $\langle\mathrm{conj}(\Psi^{1-\nu}_0)|\Psi^\nu_0\rangle=1$.  In addition, we calculate the same overlap for the first-excited state, i.e., $\langle\mathrm{conj}(\Psi^{1-\nu}_1)|\Psi^\nu_1\rangle$ to examine how the particle-hole symmetry of the energy gaps is broken, i.e., is the symmetry broken in the ground state, the excited state, or both.

\begin{figure}[b!]
\includegraphics[width=3.4in]{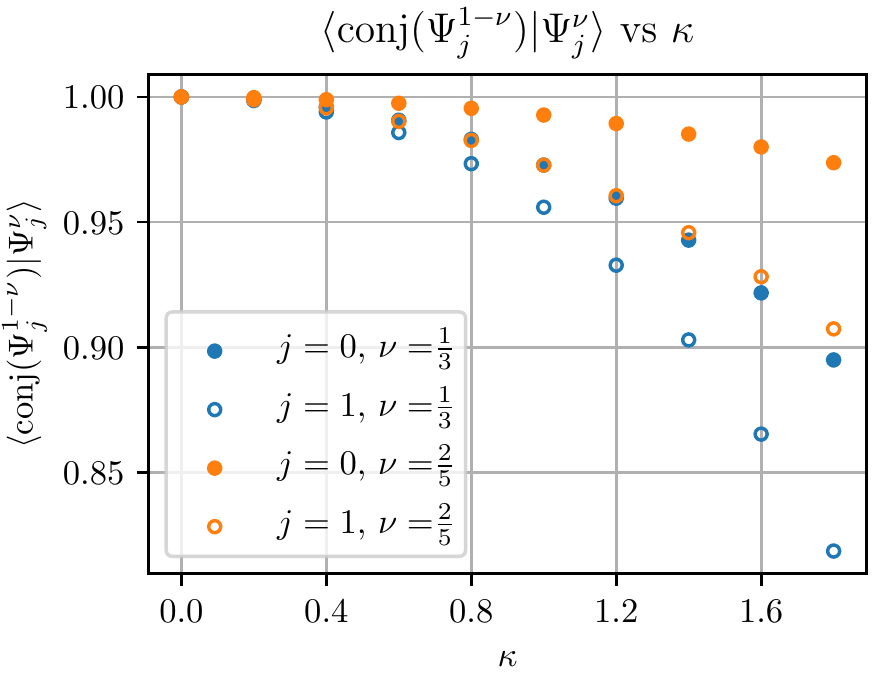}\\
\includegraphics[width=3.4in]{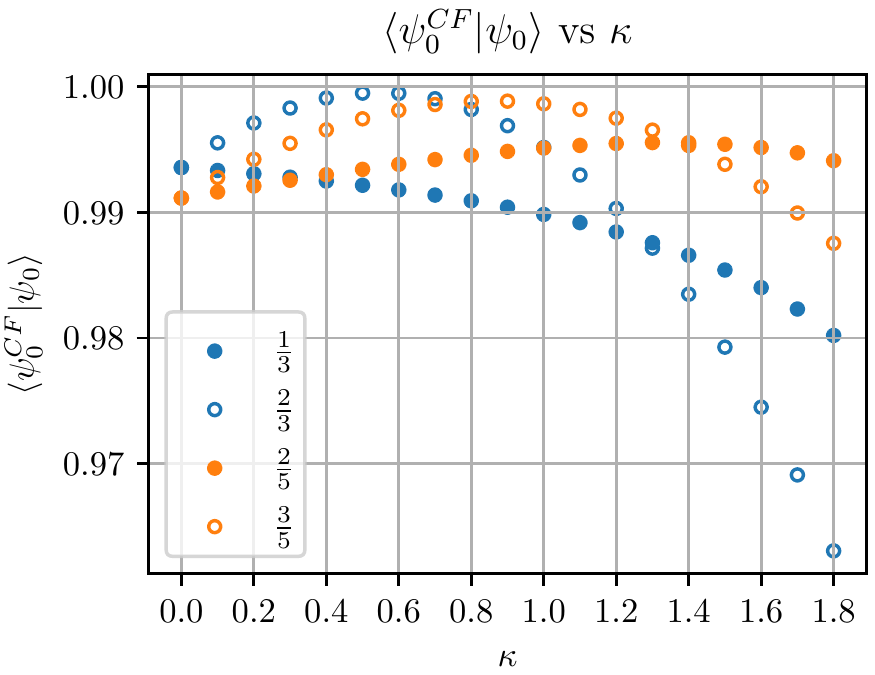}
\caption{(top) $\langle\mathrm{conj}(\Psi^{1-\nu}_j)|\Psi^\nu_j\rangle$ for $N=10(18)$, $Q=13.5$ for $\nu=\frac{1}{3}(\frac{2}{3})$ (blue symbols) and $N=12(15)$, $Q=13.0$ for $\nu=\frac{2}{5}(\frac{3}{5})$ (orange symbols) for the ground state ($j=0$, solid  symbols) and first-excited state ($j=1$, open  symbols). The overlaps decrease with $\kappa$, implying that the states are breaking particle-hole symmetry. Ground states are more resilient to the LL-mixing-driven particle-hole symmetry breaking than the first-excited states. Furthermore, the $\nu=\frac{2}{5}(\frac{3}{5})$ is more robust than $\nu=\frac{1}{3}(\frac{2}{3})$.  (bottom) The overlaps of the exact ground state with the composite fermion ground state for same systems as above. The exact ground states, even under LL mixing perturbations, are very close to the approximated composite fermion ground state.}
\label{fig:ph_sym}
\end{figure}

The top panel of Fig.~\ref{fig:ph_sym} shows $\langle\mathrm{conj}(\Psi^{1-\nu}_j)|\Psi^\nu_j\rangle$ as a function of $\kappa$ for the ground ($j=0$) and first-excited states ($j=1$). For strong magnetic fields (small $\kappa$) the overlaps are very close to unity--note the scale on the ordinate axis--the overlap is always $\approx 0.9$ or above, and for the ground state at $\nu=\frac{2}{5} (\frac{3}{5})$ it is always above approximately $0.97$. As $\kappa$ increases the overlaps monotonically decrease from unity. The excited state, however, shows more severe particle-hole symmetry breaking as measured via $\langle\mathrm{conj}(\Psi^{1-\nu}_1)|\Psi^\nu_1\rangle$ compared with the ground state.  Again, $\nu=\frac{1}{3}(\frac{2}{3})$ is broken more severely but even for large $\kappa$ the overlap is above $0.8$ at worst. Evidently, the ground state is more resilient to LL mixing, and therefore remains more particle-hole symmetric, as opposed to the first-excited state.

Finally, we  examine how closely the exact ground states at $\nu$ compare with the composite fermion wave functions.  For this calculation we  approximated the composite fermion ground states by exactly diagonalizing the so-called  ``hard-core" Hamiltonian consisting of taking the $V^{(2)}_{1}\to\infty$ for $\kappa=0$. The overlaps between the exact ground states and the composite fermion states are shown in the bottom panel of Fig~\ref{fig:ph_sym}. Overall, the overlaps are very large for all $\kappa$ and therefore the exact ground-state wave functions are extremely similar (the overlaps are above 0.97) to the composite fermion wave functions. The overlaps of the ground states for both $\nu=\frac{2}{3}$ and $\frac{3}{5}$ show that they approach the composite fermion ground state for large and finite magnetic fields; these overlaps nearly reach unity--the slight initial increase, before again decreasing, of the overlap as a function of $\kappa$ for $\nu\neq\frac{1}{3}$ is not fully understood; however, we emphasize that this increase occurs for overlaps above 0.99 and is possibly a finite-size effect. The $\nu=\frac{1}{3}$ overlap exhibits a similar behavior but is less pronounced.

Examination of various wave-function overlaps support the  finding that  the ground states for $\nu>1/2$ are more resilient to variations in $\kappa$ than the fractions $\nu<1/2$; however, there is  evidence of particle-hole symmetry breaking amongst these eigenstates, in particular, in the first-excited states.

\section{Experimental Comparison}
\label{sec:exp}
We now compare our results to recent experimental observations. As mentioned above, the composite fermion theory~\cite{jain2007composite,Halperin93} shapes our understanding of the physics of the fractional quantum Hall effect, especially in the lowest LL. It is expected that the energy gap, ultimately responsible for the plateaus in the Hall resistance, can be considered the  cyclotron gap of the composite fermion system, i.e., $\hbar\omega^*=\hbar e B^*/m^*c$, where $m^*$ is the mass of the composite fermions  dynamically generated  through  electron-electron interactions.  This mass $m^*$ is often called the activation mass~\cite{jain2007composite}. We can trade filling factor for density in the equation for the effective magnetic field experienced by the composite fermions ($B^*=B-2p\rho\phi_0$) yielding $B^*=\pm B/(2n+1)$ where the plus sign corresponds to $\nu=n/(2n+1)$ and the minus sign corresponds to $1-\nu=(n+1)/(2n+1)$.  Then,
\BEq
\label{eq:cfs}
\Delta E_\nu=\hbar\omega^*&=&\hbar \frac{eB^*}{m^*c}=\frac{\hbar e B}{m^*c}\frac{1}{2n+1}\\
&\approx& 1.34 \left(\frac{m_e}{m^*}\right)\frac{B}{2n+1}\;,
\EEq
in units of kelvin.  We have used $eB/m c\approx 20\;B$[T] K and $m\approx 0.067 m_e$ relevant to the GaAs systems studied in Ref.~\onlinecite{Pan20}.  For the particle-hole conjugate filling factor we find 
\BEq
\label{eq:cfholes}
\Delta E_{1-\nu}&\approx&-1.34\left(\frac{m_e}{m^*}\right)\frac{B}{2n+1}\;.
\EEq
To connect to the literature we write $\Delta E \approx \pm 1.34 (m_e/m^*) B/|2n+1|$ where we consider a plus for $\nu$ and minus for $1-\nu$.

Because the energy gap is in units of $e^2/\epsilon\ell_0$ it will scale with magnetic field as $\sqrt{B}$.  Thus,  $m^*$ also  scales like $\sqrt{B}$ since it is  generated through electron-electron interactions.  Following previous works, we equate our calculated neutral gaps  to the activation gap measured in experiments and plot $\Delta E$ versus $\sqrt{B}/(2n+1)$.  Noninteracting composite fermion theory predicts that $\Delta E$ will be linear in $\sqrt{B}/(2n+1)$ with a slope directly proportional to $m^*$ since $m^*$ is generated through electron-electron interactions and scales like $\sqrt{B}$. This ``V"-diagram was measured experimentally by Pan \textit{et al}.~\cite{Pan20} for $\nu=\frac{1}{3}(\frac{2}{3}),\ldots,\frac{4}{9}(\frac{5}{9})$ for magnetic-field strengths spanning $B=4.5$--$13.5$ T.  These magnetic fields correspond to a LL mixing parameter $\kappa$ ranging from $\kappa=1.19$ to $0.69$.  Experimentally it was found that the energy gap $\Delta E$ behaved approximately linear when plotted versus $\sqrt{B}/|2n+1|$ for all $\nu$ and $1-\nu$, with approximately the same slope, indicating that particle-hole symmetry was largely maintained~\cite{Pan20}.  Disorder broadening was also found to be particle-hole symmetric since gaps at $\nu$ and $1-\nu$ both extrapolated to $\Gamma \approx -1.8$ K at $\nu=1/2$.  This linearity translates to a composite fermion mass that is approximately particle-hole symmetric, independent of $\nu$, and $m^*\approx 0.2\;m_e$, modulo $\sqrt{B}$.  We note that previous calculations, without including LL mixing or finite thickness, found $m^*\approx 0.1\;m_e$\cite{Halperin93,Morf1995PRL,Morf02}--this theoretical value has been noted to be about a factor of two too small~\cite{jain2007composite}.

\begin{figure}
\includegraphics[width=3.in]{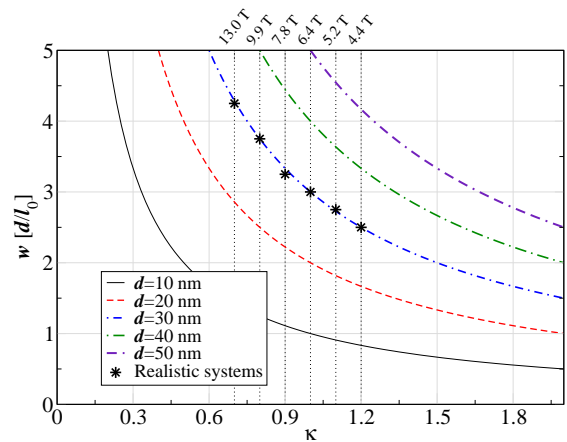}
\caption{Systems studied in the $w$--$\kappa$ plane are indicated by the stars.  The blue dashed-dotted curve corresponds to a system with a quantum well thickness of 30 nm while the other curves correspond to systems ranging from thickness of 10 nm to 50 nm.}
\label{fig:plot}
\end{figure}

\begin{table}[!h]
\caption{\label{table:systems} Realistic systems studied to compare with the experiment in Ref.~\onlinecite{Pan20}.}
\begin{ruledtabular}
\begin{tabular}{lcr}
$B$[T] & $\kappa$ & $w$  \\
\hline
4.4 & 1.2  & 2.5 \\
5.2 & 1.1  & 2.75\\
6.4 & 1.0  & 3.0\\
7.8 & 0.9  & 3.25\\
9.9 & 0.8  & 3.75\\
13.0 & 0.7 & 4.25
\end{tabular}
\end{ruledtabular}
\end{table}
The experimental systems corresponding to a constant magnetic-field strength and changing filling factor  trace out a path in the $w$--$\kappa$ plane as shown in Fig.~\ref{fig:plot}; $w$ is the width of the confining potential in the direction perpendicular to the two-dimensional plane in units of the magnetic length $\ell_0$.
To compare our calculations to Ref.~\onlinecite{Pan20} we need to consider a system with finite well width, and therefore sub-band mixing, in addition to LL mixing.  The experiment used a heterojunction insulated-gate field-effect transistor (HIGFET) in a  GaAs/AlGaAs heterostructure. This system is usually modeled via the Fang-Howard potential~\cite{FangPRL1966,AndoRMP1982,SternPRB1984} with a thickness chosen variationally.  For technical reasons, this potential is hard to include in our LL mixing calculations and we therefore consider a ``narrow"  square well potential with a thickness of $d=30$ nm chosen to approximate the thickness of the HIGFET--we note that different confining potentials can be used to model real systems with appropriately rescaled thickness parameters~\cite{Peterson08b}. Thus, we consider a series of systems with widths
\BEq
w=d/\ell_0= (30\mathrm{\;nm}/25\mathrm{\;nm})\sqrt{B}[\mathrm{T}]=1.2\sqrt{B}[\mathrm{T}]\;,
\EEq
measured in units of magnetic length, and $\kappa=2.5/\sqrt{B}[\mathrm{T}]$.  Following Ref.~\onlinecite{Peterson13b} we approximate the quantum well using an infinite square well potential and consider the systems shown in Table~\ref{table:systems}.

\begin{figure*}[t!]
\includegraphics[width=2.25in]{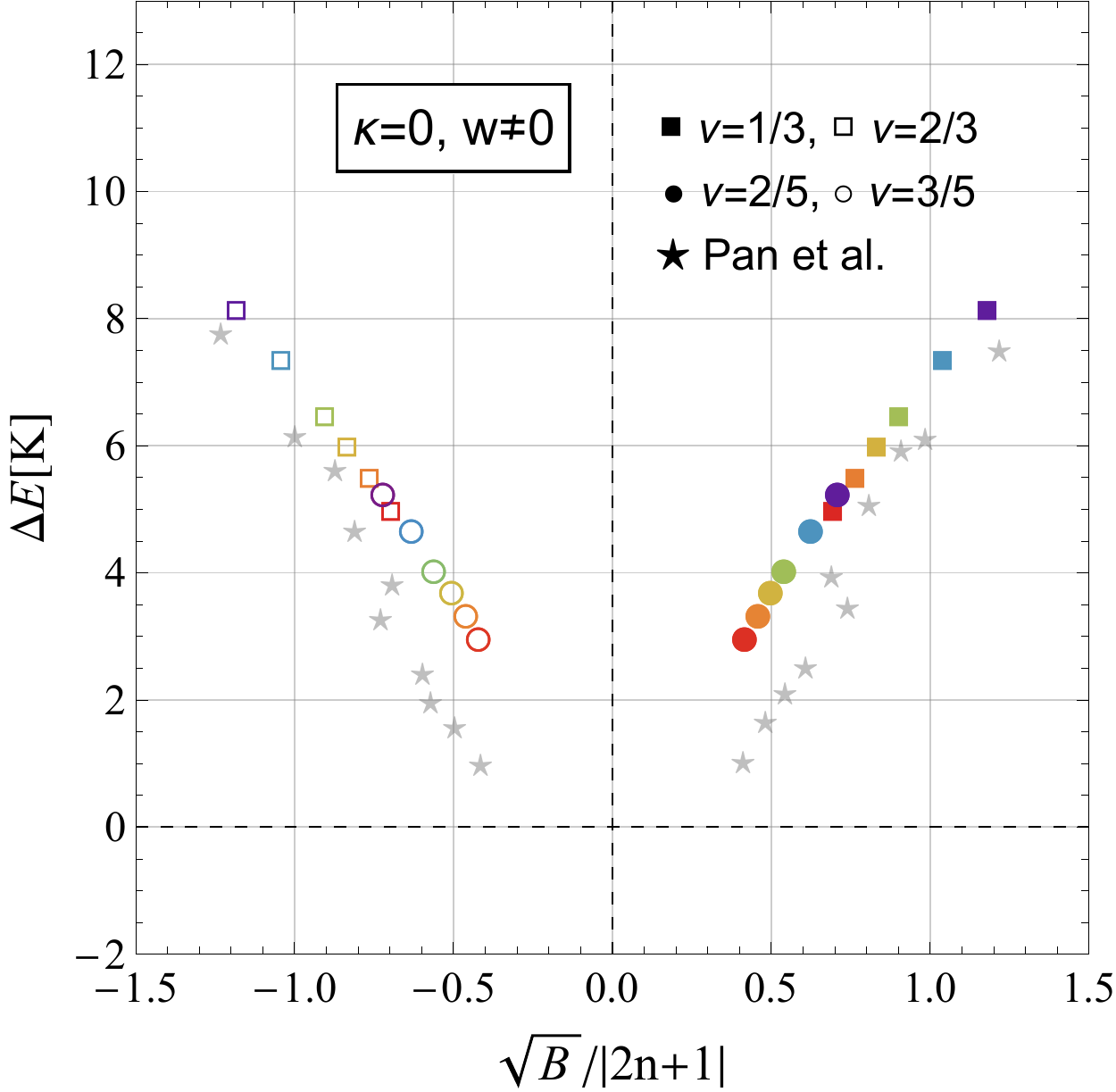}
\includegraphics[width=2.25in]{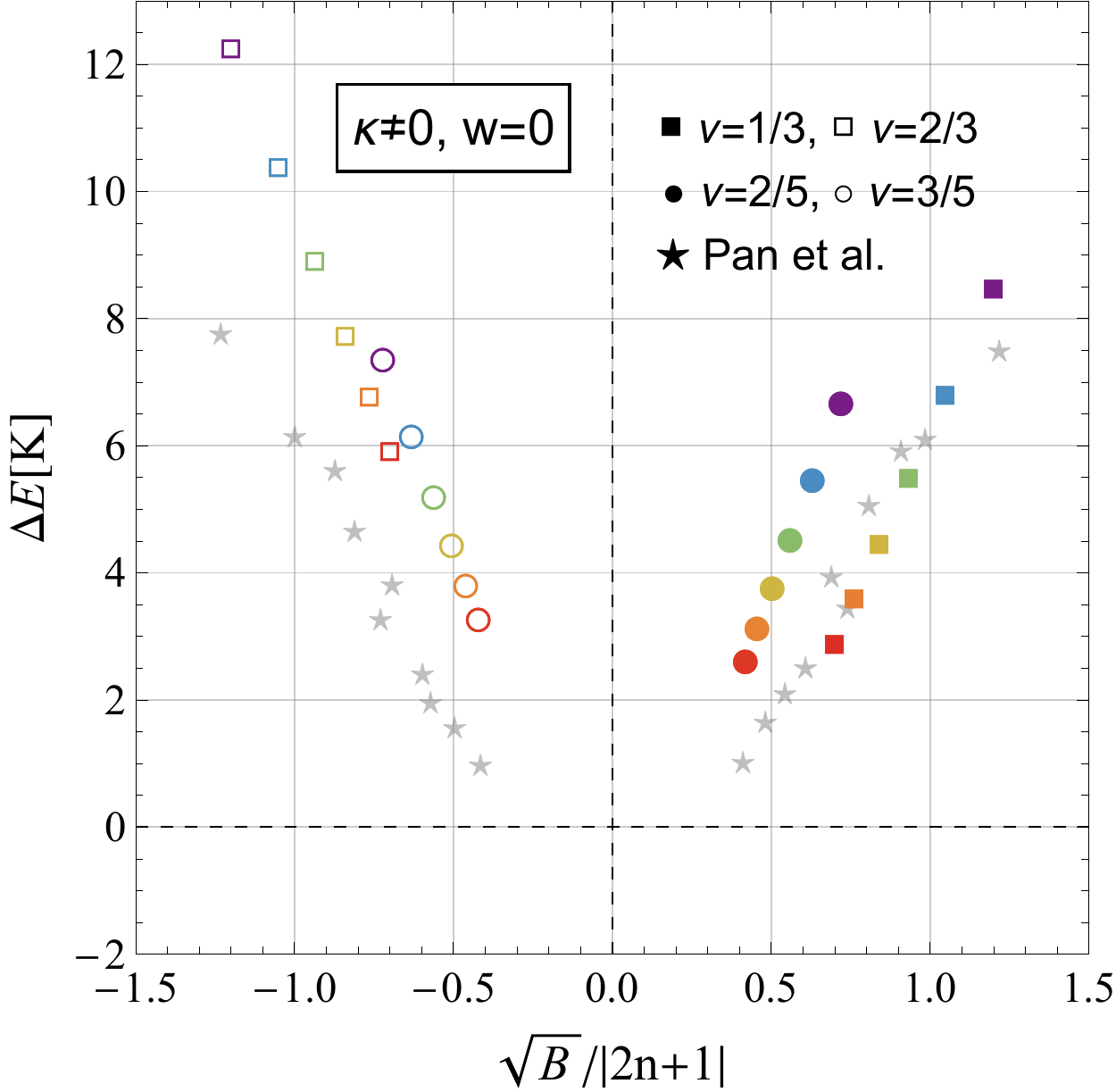}
\includegraphics[width=2.25in]{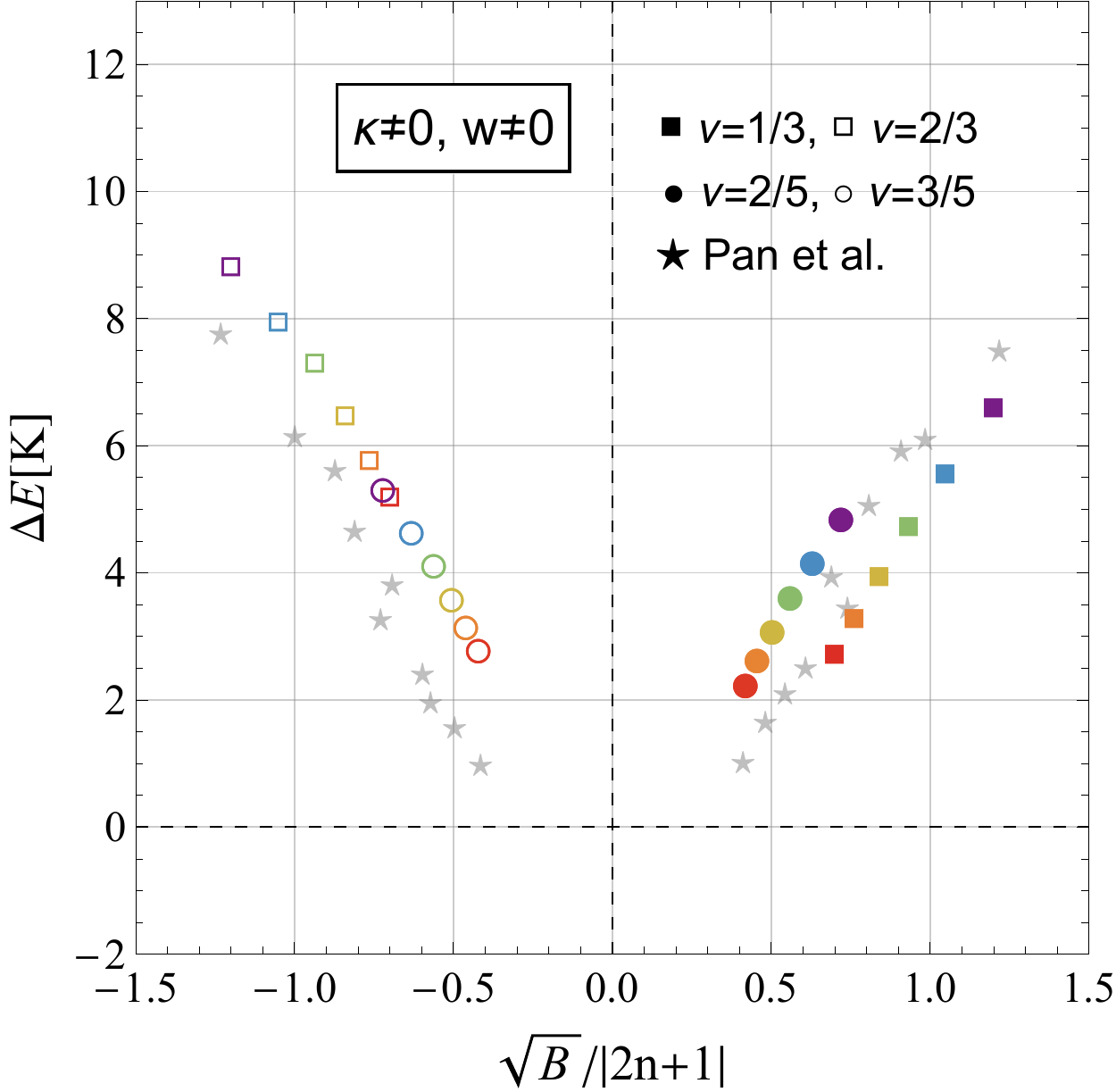}
\caption{Calculated energy gaps $\Delta E$ in units of kelvin as functions of $\sqrt{B}/|2n+1|$.  Positive $2n+1$ corresponds to $\nu<1/2$ while negative $2n+1$ corresponds to $\nu>1/2$, respectively.  The stars represent experimental points taken from Pan \textit{et al.}~\cite{Pan20}. Filled (open) squares and filled (open) circles correspond to $\nu=\frac{1}{3}$ ($\frac{2}{3}$) and $\nu=\frac{2}{5}$ ($\frac{3}{5}$), respectively, with ``cool" colors corresponding to large magnetic-field strengths ($B=13$ T, $\kappa=1.2$) and ``warm" colors to smaller magnetic-field strengths ($B=4.4$ T, $\kappa=0.7$).  The three panels (from left to right) represent situations where ($\kappa=0,\;w\neq0$), ($\kappa\neq0,\;w=0$), and ($\kappa\neq0,\;w\neq 0$), respectively, with the last case the ``realistic" case.}
\label{fig:gap_vs_B}
\end{figure*}

In Fig.~\ref{fig:gap_vs_B} we plot our theoretically calculated energy gaps $\Delta E$ as a function of $\sqrt{B}/|2n+1|$ for $B=13$--$4.4$ T corresponding to a realistic system approximating the sample of Ref.~\onlinecite{Pan20}.  We have offset the energy gaps by a disorder broadening term $\Gamma=1.8$ K used by Pan \textit{et al.}~\cite{Pan20}. In addition, we have considered three different scenarios: (i) finite thickness with no LL mixing ($\kappa=0,\;w\neq 0$), (ii) LL mixing but no thickness ($\kappa\neq0,\;w=0$), and (iii) the full ``realistic" case with both LL and sub-band mixing ($\kappa\neq 0,\;w\neq 0$).  Below we have also considered the effective mass in the situation of zero thickness and no LL and sub-band mixing  ($\kappa=0,\;w=0$).

For all three cases, the energy gap is found to be approximately the same magnitude of the experimentally measured gaps (although slightly too large, as is typical).  Interestingly, we find that finite thickness ($w\neq 0)$ is crucial in order to best capture the experimental data--finite thickness both reduces all gaps but also reduces the gaps more for larger magnetic-field strengths which reduces the slope of $\Delta E$ as a function of $\sqrt{B}/|2n+1|$ increasing the effective mass.  Of course, without LL and sub-band mixing there is no chance of particle-hole symmetry breaking.  In the ``realistic" case (right panel) we find for $\nu>1/2$ the results are largely unchanged, however, the gaps for $\nu<1/2$ are further reduced (bringing them closer to experiment) indicating that particle-hole symmetry apparently is broken in our calculations, as measured via the energy gaps. However, the $\nu=\frac{1}{3}$ system behaves almost as an outlier in that the energy gaps are reduced more than those for $\nu=\frac{2}{3}$ when LL mixing is included. The reason for this behavior is an open question and involves the effects of the complicated three-body terms.  However,  as shown below, the slope (and effective mass) of the $\nu=\frac{1}{3}$ is largely unchanged by the inclusion of LL mixing. 

By  determining the slope of each best-fit line to the data for each $\nu$ in Fig.~\ref{fig:gap_vs_B} we can extract the composite fermion mass $m^*$.  Table~\ref{table:masses} shows that the extracted masses are approximately independent of $\nu$. Furthermore, it was noted previously in the absence of LL mixing that finite width would suppress the gaps and enhance the effective mass~\cite{Park98}. We also find an enhancement of the composite fermion effective mass $m^*$ but how this happens in the realistic system is subtle. As the value of the magnetic-field strength is increased at fixed filling factor both the LL mixing parameter $\kappa$  decreases while the width $w$ of the confining potential, in units of magnetic length,  increases.  Decreasing $\kappa$ for fixed width increases the gaps towards the infinite magnetic-field limit.  Meanwhile, increasing the width $w$ for fixed $\kappa$ decreases the gaps.  The two effects somewhat offset one another (see Table~\ref{table:systems}) as can be observed in Fig.~\ref{fig:gap_vs_B} and Table~\ref{table:masses}. Finally, we find that under realistic conditions ($\kappa\neq 0$, $w\neq 0$) our effective-mass calculations are in line quantitatively with those recently measured in Ref.~\onlinecite{Pan20}.

\begin{table}[]
\caption{\label{table:masses} Composite fermion mass $m^*$, moduli $\sqrt{B}$ determined from the slope of the best-fit line of $\Delta E_\nu$ vs. $\sqrt{B}/|2n+1|$ using Eqs.~\ref{eq:cfs} and~\ref{eq:cfholes}. 
The standard error in the linear fits are in the third or fourth digit for all cases. }
\begin{ruledtabular}
\begin{tabular}{l|c|cccc}
   &  & $\nu=\frac{1}{3}$ & $\frac{2}{3}$ & $\frac{2}{5}$ & $\frac{3}{5}$ \\
\hline
$(\kappa=0,\;w=0)$ & $m^*/m_e$ & 0.12  & 0.12  & 0.10  & 0.10  \\
$(\kappa=0,\;w\neq0)$ & $m^*/m_e$ & 0.21  & 0.21  & 0.17  & 0.18  \\
$(\kappa\neq 0,\;w=0)$ & $m^*/m_e$ & 0.12  & 0.11  & 0.10  & 0.10  \\
$(\kappa\neq 0,\;w\neq 0$) & $m^*/m_e$ & 0.17  & 0.18  & 0.15  & 0.16
\end{tabular}
\end{ruledtabular}
\end{table}

\section{Conclusion}
\label{sec:conc}

In this work we have studied particle-hole symmetry breaking in the lowest Landau level of the fractional quantum Hall effect using exact diagonalization of finite systems in the spherical geometry by investigating the Hamiltonian pseudopotentials, the energy gaps, and wave-function overlaps. We find that particle-hole symmetry is broken under LL mixing as measured by the energy gaps.  However, the ground states remain largely particle-hole symmetric as measured via wave-function overlaps while the first-excited states show more particle-hole symmetry breaking.  Finally we compare our theoretical calculations to recent experimental results~\cite{Pan20} and find that while the energy gaps show modest particle-hole symmetry breaking, the effective mass of the composite fermions is largely particle-hole symmetric and our realistic effects were able to capture the measured effective mass quantitatively. 

We note that very little has been studied theoretically with regards to spin physics and LL mixing~\cite{Yuhe2016} and reiterate that the conflicting results at $\nu=\frac{1}{3}$ ($\frac{2}{3}$) between exact diagonalization (the present work) and the fixed phase diffusion Monte Carlo~\cite{Balram16} results remain an open question. 
 
\section*{Acknowledgments}
We thank Vito Scarola for helpful comments on the paper. This research was supported by the National Science Foundation under Grant No. DMR-1508290 and the Office of Research
and Sponsored Programs at California State University Long Beach.




\begin{thebibliography}{58}
\expandafter\ifx\csname natexlab\endcsname\relax\def\natexlab#1{#1}\fi
\expandafter\ifx\csname bibnamefont\endcsname\relax
  \def\bibnamefont#1{#1}\fi
\expandafter\ifx\csname bibfnamefont\endcsname\relax
  \def\bibfnamefont#1{#1}\fi
\expandafter\ifx\csname citenamefont\endcsname\relax
  \def\citenamefont#1{#1}\fi
\expandafter\ifx\csname url\endcsname\relax
  \def\url#1{\texttt{#1}}\fi
\expandafter\ifx\csname urlprefix\endcsname\relax\def\urlprefix{URL }\fi
\providecommand{\bibinfo}[2]{#2}
\providecommand{\eprint}[2][]{\url{#2}}

\bibitem[{\citenamefont{Tsui et~al.}(1982)\citenamefont{Tsui, Stormer, and
  Gossard}}]{Tsui82}
\bibinfo{author}{\bibfnamefont{D.~C.} \bibnamefont{Tsui}},
  \bibinfo{author}{\bibfnamefont{H.~L.} \bibnamefont{Stormer}},
  \bibnamefont{and} \bibinfo{author}{\bibfnamefont{A.~C.}
  \bibnamefont{Gossard}}, \bibinfo{journal}{Phys. Rev. Lett.}
  \textbf{\bibinfo{volume}{48}}, \bibinfo{pages}{1559} (\bibinfo{year}{1982}),
  \urlprefix\url{https://link.aps.org/doi/10.1103/PhysRevLett.48.1559}.

\bibitem[{Pra()}]{Prange90}
\bibinfo{note}{\textit{The Quantum Hall Effect} edited by R. Prange and S. M.
  Girvin (Springer-Verlag, New York, 1990)}.

\bibitem[{Das()}]{DasSarma97}
\bibinfo{note}{\textit{Perspectives in Quantum Hall Effects: novel quantum
  liquids in low-dimensional semiconductor structures} edited by S. Das Sarma
  and A. Pinczuk (Wiley, New York, 1997)}.

\bibitem[{jai()}]{jain2007composite}
\bibinfo{note}{J. K. Jain, \textit{Composite Fermions} (Cambridge University
  Press, Cambridge, 2007)}.

\bibitem[{\citenamefont{Kitaev}(2003)}]{Kitaev97}
\bibinfo{author}{\bibfnamefont{A.}~\bibnamefont{Kitaev}},
  \bibinfo{journal}{Annals of Physics} \textbf{\bibinfo{volume}{303}},
  \bibinfo{pages}{2} (\bibinfo{year}{2003}), ISSN \bibinfo{issn}{0003-4916},
  \urlprefix\url{https://www.sciencedirect.com/science/article/pii/S0003491602000180}.

\bibitem[{\citenamefont{Freedman}(2001)}]{Freedman01}
\bibinfo{author}{\bibfnamefont{M.~H.} \bibnamefont{Freedman}},
  \bibinfo{journal}{Foundations of Computational Mathematics}
  \textbf{\bibinfo{volume}{1}}, \bibinfo{pages}{183} (\bibinfo{year}{2001}),
  \urlprefix\url{https://doi.org/10.1007/s102080010006}.

\bibitem[{\citenamefont{Nayak et~al.}(2008)\citenamefont{Nayak, Simon, Stern,
  Freedman, and Das~Sarma}}]{Nayak08}
\bibinfo{author}{\bibfnamefont{C.}~\bibnamefont{Nayak}},
  \bibinfo{author}{\bibfnamefont{S.~H.} \bibnamefont{Simon}},
  \bibinfo{author}{\bibfnamefont{A.}~\bibnamefont{Stern}},
  \bibinfo{author}{\bibfnamefont{M.}~\bibnamefont{Freedman}}, \bibnamefont{and}
  \bibinfo{author}{\bibfnamefont{S.}~\bibnamefont{Das~Sarma}},
  \bibinfo{journal}{Rev. Mod. Phys.} \textbf{\bibinfo{volume}{80}},
  \bibinfo{pages}{1083} (\bibinfo{year}{2008}),
  \urlprefix\url{https://link.aps.org/doi/10.1103/RevModPhys.80.1083}.

\bibitem[{\citenamefont{Willett et~al.}(1987)\citenamefont{Willett, Eisenstein,
  St\"ormer, Tsui, Gossard, and English}}]{Willett87}
\bibinfo{author}{\bibfnamefont{R.}~\bibnamefont{Willett}},
  \bibinfo{author}{\bibfnamefont{J.~P.} \bibnamefont{Eisenstein}},
  \bibinfo{author}{\bibfnamefont{H.~L.} \bibnamefont{St\"ormer}},
  \bibinfo{author}{\bibfnamefont{D.~C.} \bibnamefont{Tsui}},
  \bibinfo{author}{\bibfnamefont{A.~C.} \bibnamefont{Gossard}},
  \bibnamefont{and} \bibinfo{author}{\bibfnamefont{J.~H.}
  \bibnamefont{English}}, \bibinfo{journal}{Phys. Rev. Lett.}
  \textbf{\bibinfo{volume}{59}}, \bibinfo{pages}{1776} (\bibinfo{year}{1987}),
  \urlprefix\url{https://link.aps.org/doi/10.1103/PhysRevLett.59.1776}.

\bibitem[{\citenamefont{Xia et~al.}(2004)\citenamefont{Xia, Pan, Vicente,
  Adams, Sullivan, Stormer, Tsui, Pfeiffer, Baldwin, and West}}]{Xia04}
\bibinfo{author}{\bibfnamefont{J.~S.} \bibnamefont{Xia}},
  \bibinfo{author}{\bibfnamefont{W.}~\bibnamefont{Pan}},
  \bibinfo{author}{\bibfnamefont{C.~L.} \bibnamefont{Vicente}},
  \bibinfo{author}{\bibfnamefont{E.~D.} \bibnamefont{Adams}},
  \bibinfo{author}{\bibfnamefont{N.~S.} \bibnamefont{Sullivan}},
  \bibinfo{author}{\bibfnamefont{H.~L.} \bibnamefont{Stormer}},
  \bibinfo{author}{\bibfnamefont{D.~C.} \bibnamefont{Tsui}},
  \bibinfo{author}{\bibfnamefont{L.~N.} \bibnamefont{Pfeiffer}},
  \bibinfo{author}{\bibfnamefont{K.~W.} \bibnamefont{Baldwin}},
  \bibnamefont{and} \bibinfo{author}{\bibfnamefont{K.~W.} \bibnamefont{West}},
  \bibinfo{journal}{Phys. Rev. Lett.} \textbf{\bibinfo{volume}{93}},
  \bibinfo{pages}{176809} (\bibinfo{year}{2004}),
  \urlprefix\url{https://link.aps.org/doi/10.1103/PhysRevLett.93.176809}.

\bibitem[{\citenamefont{Bishara and Nayak}(2009)}]{Bishara09a}
\bibinfo{author}{\bibfnamefont{W.}~\bibnamefont{Bishara}} \bibnamefont{and}
  \bibinfo{author}{\bibfnamefont{C.}~\bibnamefont{Nayak}},
  \bibinfo{journal}{Phys. Rev. B} \textbf{\bibinfo{volume}{80}},
  \bibinfo{pages}{121302(R)} (\bibinfo{year}{2009}),
  \urlprefix\url{https://link.aps.org/doi/10.1103/PhysRevB.80.121302}.

\bibitem[{\citenamefont{Peterson and Nayak}(2013)}]{Peterson13b}
\bibinfo{author}{\bibfnamefont{M.~R.} \bibnamefont{Peterson}} \bibnamefont{and}
  \bibinfo{author}{\bibfnamefont{C.}~\bibnamefont{Nayak}},
  \bibinfo{journal}{Phys. Rev. B} \textbf{\bibinfo{volume}{87}},
  \bibinfo{pages}{245129} (\bibinfo{year}{2013}),
  \urlprefix\url{http://link.aps.org/doi/10.1103/PhysRevB.87.245129}.

\bibitem[{\citenamefont{Sodemann and MacDonald}(2013)}]{Sodemann13}
\bibinfo{author}{\bibfnamefont{I.}~\bibnamefont{Sodemann}} \bibnamefont{and}
  \bibinfo{author}{\bibfnamefont{A.~H.} \bibnamefont{MacDonald}},
  \bibinfo{journal}{Phys. Rev. B} \textbf{\bibinfo{volume}{87}},
  \bibinfo{pages}{245425} (\bibinfo{year}{2013}),
  \urlprefix\url{https://link.aps.org/doi/10.1103/PhysRevB.87.245425}.

\bibitem[{\citenamefont{Simon and Rezayi}(2013)}]{Simon13}
\bibinfo{author}{\bibfnamefont{S.~H.} \bibnamefont{Simon}} \bibnamefont{and}
  \bibinfo{author}{\bibfnamefont{E.~H.} \bibnamefont{Rezayi}},
  \bibinfo{journal}{Phys. Rev. B} \textbf{\bibinfo{volume}{87}},
  \bibinfo{pages}{155426} (\bibinfo{year}{2013}),
  \urlprefix\url{https://link.aps.org/doi/10.1103/PhysRevB.87.155426}.

\bibitem[{\citenamefont{Wooten et~al.}(2013)\citenamefont{Wooten, Macek, and
  Quinn}}]{Wooten13}
\bibinfo{author}{\bibfnamefont{R.~E.} \bibnamefont{Wooten}},
  \bibinfo{author}{\bibfnamefont{J.~H.} \bibnamefont{Macek}}, \bibnamefont{and}
  \bibinfo{author}{\bibfnamefont{J.~J.} \bibnamefont{Quinn}},
  \bibinfo{journal}{Phys. Rev. B} \textbf{\bibinfo{volume}{88}},
  \bibinfo{pages}{155421} (\bibinfo{year}{2013}),
  \urlprefix\url{https://link.aps.org/doi/10.1103/PhysRevB.88.155421}.

\bibitem[{\citenamefont{W\'ojs et~al.}(2010)\citenamefont{W\'ojs,
  T\ifmmode~\mbox{\H{o}}\else \H{o}\fi{}ke, and Jain}}]{Wojs10}
\bibinfo{author}{\bibfnamefont{A.}~\bibnamefont{W\'ojs}},
  \bibinfo{author}{\bibfnamefont{C.}~\bibnamefont{T\ifmmode~\mbox{\H{o}}\else
  \H{o}\fi{}ke}}, \bibnamefont{and} \bibinfo{author}{\bibfnamefont{J.~K.}
  \bibnamefont{Jain}}, \bibinfo{journal}{Phys. Rev. Lett.}
  \textbf{\bibinfo{volume}{105}}, \bibinfo{pages}{096802}
  (\bibinfo{year}{2010}),
  \urlprefix\url{https://link.aps.org/doi/10.1103/PhysRevLett.105.096802}.

\bibitem[{\citenamefont{Rezayi and Simon}(2011)}]{Rezayi11}
\bibinfo{author}{\bibfnamefont{E.~H.} \bibnamefont{Rezayi}} \bibnamefont{and}
  \bibinfo{author}{\bibfnamefont{S.~H.} \bibnamefont{Simon}},
  \bibinfo{journal}{Phys. Rev. Lett.} \textbf{\bibinfo{volume}{106}},
  \bibinfo{pages}{116801} (\bibinfo{year}{2011}),
  \urlprefix\url{https://link.aps.org/doi/10.1103/PhysRevLett.106.116801}.

\bibitem[{\citenamefont{Son}(2015)}]{Son15}
\bibinfo{author}{\bibfnamefont{D.~T.} \bibnamefont{Son}},
  \bibinfo{journal}{Phys. Rev. X} \textbf{\bibinfo{volume}{5}},
  \bibinfo{pages}{031027} (\bibinfo{year}{2015}),
  \urlprefix\url{https://link.aps.org/doi/10.1103/PhysRevX.5.031027}.

\bibitem[{\citenamefont{Peterson et~al.}(2015)\citenamefont{Peterson, Wu,
  Cheng, Barkeshli, Wang, and Das~Sarma}}]{PetersonPRB2015}
\bibinfo{author}{\bibfnamefont{M.~R.} \bibnamefont{Peterson}},
  \bibinfo{author}{\bibfnamefont{Y.-L.} \bibnamefont{Wu}},
  \bibinfo{author}{\bibfnamefont{M.}~\bibnamefont{Cheng}},
  \bibinfo{author}{\bibfnamefont{M.}~\bibnamefont{Barkeshli}},
  \bibinfo{author}{\bibfnamefont{Z.}~\bibnamefont{Wang}}, \bibnamefont{and}
  \bibinfo{author}{\bibfnamefont{S.}~\bibnamefont{Das~Sarma}},
  \bibinfo{journal}{Phys. Rev. B} \textbf{\bibinfo{volume}{92}},
  \bibinfo{pages}{035103} (\bibinfo{year}{2015}),
  \urlprefix\url{https://link.aps.org/doi/10.1103/PhysRevB.92.035103}.

\bibitem[{\citenamefont{Pakrouski et~al.}(2015)\citenamefont{Pakrouski,
  Peterson, Jolicoeur, Scarola, Nayak, and Troyer}}]{Pakrouski15}
\bibinfo{author}{\bibfnamefont{K.}~\bibnamefont{Pakrouski}},
  \bibinfo{author}{\bibfnamefont{M.~R.} \bibnamefont{Peterson}},
  \bibinfo{author}{\bibfnamefont{T.}~\bibnamefont{Jolicoeur}},
  \bibinfo{author}{\bibfnamefont{V.~W.} \bibnamefont{Scarola}},
  \bibinfo{author}{\bibfnamefont{C.}~\bibnamefont{Nayak}}, \bibnamefont{and}
  \bibinfo{author}{\bibfnamefont{M.}~\bibnamefont{Troyer}},
  \bibinfo{journal}{Phys. Rev. X} \textbf{\bibinfo{volume}{5}},
  \bibinfo{pages}{021004} (\bibinfo{year}{2015}),
  \urlprefix\url{https://link.aps.org/doi/10.1103/PhysRevX.5.021004}.

\bibitem[{\citenamefont{Zaletel et~al.}(2015)\citenamefont{Zaletel, Mong,
  Pollmann, and Rezayi}}]{Zaletel15}
\bibinfo{author}{\bibfnamefont{M.~P.} \bibnamefont{Zaletel}},
  \bibinfo{author}{\bibfnamefont{R.~S.~K.} \bibnamefont{Mong}},
  \bibinfo{author}{\bibfnamefont{F.}~\bibnamefont{Pollmann}}, \bibnamefont{and}
  \bibinfo{author}{\bibfnamefont{E.~H.} \bibnamefont{Rezayi}},
  \bibinfo{journal}{Phys. Rev. B} \textbf{\bibinfo{volume}{91}},
  \bibinfo{pages}{045115} (\bibinfo{year}{2015}),
  \urlprefix\url{https://link.aps.org/doi/10.1103/PhysRevB.91.045115}.

\bibitem[{\citenamefont{Pakrouski et~al.}(2016)\citenamefont{Pakrouski, Troyer,
  Wu, Das~Sarma, and Peterson}}]{Pakrouski16}
\bibinfo{author}{\bibfnamefont{K.}~\bibnamefont{Pakrouski}},
  \bibinfo{author}{\bibfnamefont{M.}~\bibnamefont{Troyer}},
  \bibinfo{author}{\bibfnamefont{Y.-L.} \bibnamefont{Wu}},
  \bibinfo{author}{\bibfnamefont{S.}~\bibnamefont{Das~Sarma}},
  \bibnamefont{and} \bibinfo{author}{\bibfnamefont{M.~R.}
  \bibnamefont{Peterson}}, \bibinfo{journal}{Phys. Rev. B}
  \textbf{\bibinfo{volume}{94}}, \bibinfo{pages}{075108}
  (\bibinfo{year}{2016}),
  \urlprefix\url{https://link.aps.org/doi/10.1103/PhysRevB.94.075108}.

\bibitem[{\citenamefont{Rezayi}(2017)}]{Rezayi17}
\bibinfo{author}{\bibfnamefont{E.~H.} \bibnamefont{Rezayi}},
  \bibinfo{journal}{Phys. Rev. Lett.} \textbf{\bibinfo{volume}{119}},
  \bibinfo{pages}{026801} (\bibinfo{year}{2017}),
  \urlprefix\url{https://link.aps.org/doi/10.1103/PhysRevLett.119.026801}.

\bibitem[{\citenamefont{Schreiber et~al.}(2018)\citenamefont{Schreiber,
  Samkharadze, Gardner, Lyanda-Geller, Manfra, Pfeiffer, West, and
  Cs{\'a}thy}}]{Schreiber18}
\bibinfo{author}{\bibfnamefont{K.~A.} \bibnamefont{Schreiber}},
  \bibinfo{author}{\bibfnamefont{N.}~\bibnamefont{Samkharadze}},
  \bibinfo{author}{\bibfnamefont{G.~C.} \bibnamefont{Gardner}},
  \bibinfo{author}{\bibfnamefont{Y.}~\bibnamefont{Lyanda-Geller}},
  \bibinfo{author}{\bibfnamefont{M.~J.} \bibnamefont{Manfra}},
  \bibinfo{author}{\bibfnamefont{L.~N.} \bibnamefont{Pfeiffer}},
  \bibinfo{author}{\bibfnamefont{K.~W.} \bibnamefont{West}}, \bibnamefont{and}
  \bibinfo{author}{\bibfnamefont{G.~A.} \bibnamefont{Cs{\'a}thy}},
  \bibinfo{journal}{Nature Communications} \textbf{\bibinfo{volume}{9}},
  \bibinfo{pages}{2400} (\bibinfo{year}{2018}),
  \urlprefix\url{https://doi.org/10.1038/s41467-018-04879-1}.

\bibitem[{\citenamefont{Banerjee et~al.}(2018)\citenamefont{Banerjee, Heiblum,
  Umansky, Feldman, Oreg, and Stern}}]{Banerjee18}
\bibinfo{author}{\bibfnamefont{M.}~\bibnamefont{Banerjee}},
  \bibinfo{author}{\bibfnamefont{M.}~\bibnamefont{Heiblum}},
  \bibinfo{author}{\bibfnamefont{V.}~\bibnamefont{Umansky}},
  \bibinfo{author}{\bibfnamefont{D.~E.} \bibnamefont{Feldman}},
  \bibinfo{author}{\bibfnamefont{Y.}~\bibnamefont{Oreg}}, \bibnamefont{and}
  \bibinfo{author}{\bibfnamefont{A.}~\bibnamefont{Stern}},
  \bibinfo{journal}{Nature} \textbf{\bibinfo{volume}{559}},
  \bibinfo{pages}{205} (\bibinfo{year}{2018}),
  \urlprefix\url{https://doi.org/10.1038/s41586-018-0184-1}.

\bibitem[{\citenamefont{Mross et~al.}(2018)\citenamefont{Mross, Oreg, Stern,
  Margalit, and Heiblum}}]{Mross18}
\bibinfo{author}{\bibfnamefont{D.~F.} \bibnamefont{Mross}},
  \bibinfo{author}{\bibfnamefont{Y.}~\bibnamefont{Oreg}},
  \bibinfo{author}{\bibfnamefont{A.}~\bibnamefont{Stern}},
  \bibinfo{author}{\bibfnamefont{G.}~\bibnamefont{Margalit}}, \bibnamefont{and}
  \bibinfo{author}{\bibfnamefont{M.}~\bibnamefont{Heiblum}},
  \bibinfo{journal}{Phys. Rev. Lett.} \textbf{\bibinfo{volume}{121}},
  \bibinfo{pages}{026801} (\bibinfo{year}{2018}),
  \urlprefix\url{https://link.aps.org/doi/10.1103/PhysRevLett.121.026801}.

\bibitem[{\citenamefont{Wang et~al.}(2018)\citenamefont{Wang, Vishwanath, and
  Halperin}}]{Wang18}
\bibinfo{author}{\bibfnamefont{C.}~\bibnamefont{Wang}},
  \bibinfo{author}{\bibfnamefont{A.}~\bibnamefont{Vishwanath}},
  \bibnamefont{and} \bibinfo{author}{\bibfnamefont{B.~I.}
  \bibnamefont{Halperin}}, \bibinfo{journal}{Phys. Rev. B}
  \textbf{\bibinfo{volume}{98}}, \bibinfo{pages}{045112}
  (\bibinfo{year}{2018}),
  \urlprefix\url{https://link.aps.org/doi/10.1103/PhysRevB.98.045112}.

\bibitem[{\citenamefont{Lian and Wang}(2018)}]{Lian18}
\bibinfo{author}{\bibfnamefont{B.}~\bibnamefont{Lian}} \bibnamefont{and}
  \bibinfo{author}{\bibfnamefont{J.}~\bibnamefont{Wang}},
  \bibinfo{journal}{Phys. Rev. B} \textbf{\bibinfo{volume}{97}},
  \bibinfo{pages}{165124} (\bibinfo{year}{2018}),
  \urlprefix\url{https://link.aps.org/doi/10.1103/PhysRevB.97.165124}.

\bibitem[{\citenamefont{Simon}(2018)}]{Simon18}
\bibinfo{author}{\bibfnamefont{S.~H.} \bibnamefont{Simon}},
  \bibinfo{journal}{Phys. Rev. B} \textbf{\bibinfo{volume}{97}},
  \bibinfo{pages}{121406(R)} (\bibinfo{year}{2018}),
  \urlprefix\url{https://link.aps.org/doi/10.1103/PhysRevB.97.121406}.

\bibitem[{\citenamefont{Ma and Feldman}(2019)}]{Ma19}
\bibinfo{author}{\bibfnamefont{K.~K.~W.} \bibnamefont{Ma}} \bibnamefont{and}
  \bibinfo{author}{\bibfnamefont{D.~E.} \bibnamefont{Feldman}},
  \bibinfo{journal}{Phys. Rev. B} \textbf{\bibinfo{volume}{99}},
  \bibinfo{pages}{085309} (\bibinfo{year}{2019}),
  \urlprefix\url{https://link.aps.org/doi/10.1103/PhysRevB.99.085309}.

\bibitem[{\citenamefont{Zhu and Sheng}(2019)}]{Zhu19}
\bibinfo{author}{\bibfnamefont{W.}~\bibnamefont{Zhu}} \bibnamefont{and}
  \bibinfo{author}{\bibfnamefont{D.~N.} \bibnamefont{Sheng}},
  \bibinfo{journal}{Phys. Rev. Lett.} \textbf{\bibinfo{volume}{123}},
  \bibinfo{pages}{056804} (\bibinfo{year}{2019}),
  \urlprefix\url{https://link.aps.org/doi/10.1103/PhysRevLett.123.056804}.

\bibitem[{\citenamefont{Simon et~al.}(2020{\natexlab{a}})\citenamefont{Simon,
  Ippoliti, Zaletel, and Rezayi}}]{Simon20}
\bibinfo{author}{\bibfnamefont{S.~H.} \bibnamefont{Simon}},
  \bibinfo{author}{\bibfnamefont{M.}~\bibnamefont{Ippoliti}},
  \bibinfo{author}{\bibfnamefont{M.~P.} \bibnamefont{Zaletel}},
  \bibnamefont{and} \bibinfo{author}{\bibfnamefont{E.~H.}
  \bibnamefont{Rezayi}}, \bibinfo{journal}{Phys. Rev. B}
  \textbf{\bibinfo{volume}{101}}, \bibinfo{pages}{041302(R)}
  (\bibinfo{year}{2020}{\natexlab{a}}),
  \urlprefix\url{https://link.aps.org/doi/10.1103/PhysRevB.101.041302}.


\bibitem[{\citenamefont{Levin et~al.}(2007)\citenamefont{Levin, Halperin, and
  Rosenow}}]{Levin07}
\bibinfo{author}{\bibfnamefont{M.}~\bibnamefont{Levin}},
  \bibinfo{author}{\bibfnamefont{B.~I.} \bibnamefont{Halperin}},
  \bibnamefont{and} \bibinfo{author}{\bibfnamefont{B.}~\bibnamefont{Rosenow}},
  \bibinfo{journal}{Phys. Rev. Lett.} \textbf{\bibinfo{volume}{99}},
  \bibinfo{pages}{236806} (\bibinfo{year}{2007}),
  \urlprefix\url{https://link.aps.org/doi/10.1103/PhysRevLett.99.236806}.

\bibitem[{\citenamefont{Lee et~al.}(2007)\citenamefont{Lee, Ryu, Nayak, and
  Fisher}}]{Lee07}
\bibinfo{author}{\bibfnamefont{S.-S.} \bibnamefont{Lee}},
  \bibinfo{author}{\bibfnamefont{S.}~\bibnamefont{Ryu}},
  \bibinfo{author}{\bibfnamefont{C.}~\bibnamefont{Nayak}}, \bibnamefont{and}
  \bibinfo{author}{\bibfnamefont{M.~P.~A.} \bibnamefont{Fisher}},
  \bibinfo{journal}{Phys. Rev. Lett.} \textbf{\bibinfo{volume}{99}},
  \bibinfo{pages}{236807} (\bibinfo{year}{2007}),
  \urlprefix\url{https://link.aps.org/doi/10.1103/PhysRevLett.99.236807}.

\bibitem[{\citenamefont{Moore and Read}(1991)}]{Moore91}
\bibinfo{author}{\bibfnamefont{G.}~\bibnamefont{Moore}} \bibnamefont{and}
  \bibinfo{author}{\bibfnamefont{N.}~\bibnamefont{Read}},
  \bibinfo{journal}{Nuclear Physics B} \textbf{\bibinfo{volume}{360}},
  \bibinfo{pages}{362} (\bibinfo{year}{1991}), ISSN \bibinfo{issn}{0550-3213},
  \urlprefix\url{https://www.sciencedirect.com/science/article/pii/055032139190407O}.

\bibitem[{\citenamefont{Zhu et~al.}(2015)\citenamefont{Zhu, Gong, Haldane, and
  Sheng}}]{Zhu15}
\bibinfo{author}{\bibfnamefont{W.}~\bibnamefont{Zhu}},
  \bibinfo{author}{\bibfnamefont{S.~S.} \bibnamefont{Gong}},
  \bibinfo{author}{\bibfnamefont{F.~D.~M.} \bibnamefont{Haldane}},
  \bibnamefont{and} \bibinfo{author}{\bibfnamefont{D.~N.} \bibnamefont{Sheng}},
  \bibinfo{journal}{Phys. Rev. Lett.} \textbf{\bibinfo{volume}{115}},
  \bibinfo{pages}{126805} (\bibinfo{year}{2015}),
  \urlprefix\url{https://link.aps.org/doi/10.1103/PhysRevLett.115.126805}.

\bibitem[{\citenamefont{Mong et~al.}(2017)\citenamefont{Mong, Zaletel,
  Pollmann, and Papi\ifmmode~\acute{c}\else \'{c}\fi{}}}]{Mong17}
\bibinfo{author}{\bibfnamefont{R.~S.~K.} \bibnamefont{Mong}},
  \bibinfo{author}{\bibfnamefont{M.~P.} \bibnamefont{Zaletel}},
  \bibinfo{author}{\bibfnamefont{F.}~\bibnamefont{Pollmann}}, \bibnamefont{and}
  \bibinfo{author}{\bibfnamefont{Z.}~\bibnamefont{Papi\ifmmode~\acute{c}\else
  \'{c}\fi{}}}, \bibinfo{journal}{Phys. Rev. B} \textbf{\bibinfo{volume}{95}},
  \bibinfo{pages}{115136} (\bibinfo{year}{2017}),
  \urlprefix\url{https://link.aps.org/doi/10.1103/PhysRevB.95.115136}.

\bibitem[{\citenamefont{Kumar et~al.}(2010)\citenamefont{Kumar, Cs\'athy,
  Manfra, Pfeiffer, and West}}]{Kumar10}
\bibinfo{author}{\bibfnamefont{A.}~\bibnamefont{Kumar}},
  \bibinfo{author}{\bibfnamefont{G.~A.} \bibnamefont{Cs\'athy}},
  \bibinfo{author}{\bibfnamefont{M.~J.} \bibnamefont{Manfra}},
  \bibinfo{author}{\bibfnamefont{L.~N.} \bibnamefont{Pfeiffer}},
  \bibnamefont{and} \bibinfo{author}{\bibfnamefont{K.~W.} \bibnamefont{West}},
  \bibinfo{journal}{Phys. Rev. Lett.} \textbf{\bibinfo{volume}{105}},
  \bibinfo{pages}{246808} (\bibinfo{year}{2010}),
  \urlprefix\url{https://link.aps.org/doi/10.1103/PhysRevLett.105.246808}.

\bibitem[{\citenamefont{Choi et~al.}(2008)\citenamefont{Choi, Kang, Das~Sarma,
  Pfeiffer, and West}}]{Choi08}
\bibinfo{author}{\bibfnamefont{H.~C.} \bibnamefont{Choi}},
  \bibinfo{author}{\bibfnamefont{W.}~\bibnamefont{Kang}},
  \bibinfo{author}{\bibfnamefont{S.}~\bibnamefont{Das~Sarma}},
  \bibinfo{author}{\bibfnamefont{L.~N.} \bibnamefont{Pfeiffer}},
  \bibnamefont{and} \bibinfo{author}{\bibfnamefont{K.~W.} \bibnamefont{West}},
  \bibinfo{journal}{Phys. Rev. B} \textbf{\bibinfo{volume}{77}},
  \bibinfo{pages}{081301(R)} (\bibinfo{year}{2008}),
  \urlprefix\url{https://link.aps.org/doi/10.1103/PhysRevB.77.081301}.

\bibitem[{\citenamefont{Pan et~al.}(2008)\citenamefont{Pan, Xia, Stormer, Tsui,
  Vicente, Adams, Sullivan, Pfeiffer, Baldwin, and West}}]{Pan08}
\bibinfo{author}{\bibfnamefont{W.}~\bibnamefont{Pan}},
  \bibinfo{author}{\bibfnamefont{J.~S.} \bibnamefont{Xia}},
  \bibinfo{author}{\bibfnamefont{H.~L.} \bibnamefont{Stormer}},
  \bibinfo{author}{\bibfnamefont{D.~C.} \bibnamefont{Tsui}},
  \bibinfo{author}{\bibfnamefont{C.}~\bibnamefont{Vicente}},
  \bibinfo{author}{\bibfnamefont{E.~D.} \bibnamefont{Adams}},
  \bibinfo{author}{\bibfnamefont{N.~S.} \bibnamefont{Sullivan}},
  \bibinfo{author}{\bibfnamefont{L.~N.} \bibnamefont{Pfeiffer}},
  \bibinfo{author}{\bibfnamefont{K.~W.} \bibnamefont{Baldwin}},
  \bibnamefont{and} \bibinfo{author}{\bibfnamefont{K.~W.} \bibnamefont{West}},
  \bibinfo{journal}{Phys. Rev. B} \textbf{\bibinfo{volume}{77}},
  \bibinfo{pages}{075307} (\bibinfo{year}{2008}),
  \urlprefix\url{https://link.aps.org/doi/10.1103/PhysRevB.77.075307}.

\bibitem[{\citenamefont{Zhang et~al.}(2012)\citenamefont{Zhang, Huan, Xia,
  Sullivan, Pan, Baldwin, West, Pfeiffer, and Tsui}}]{Zhang12}
\bibinfo{author}{\bibfnamefont{C.}~\bibnamefont{Zhang}},
  \bibinfo{author}{\bibfnamefont{C.}~\bibnamefont{Huan}},
  \bibinfo{author}{\bibfnamefont{J.~S.} \bibnamefont{Xia}},
  \bibinfo{author}{\bibfnamefont{N.~S.} \bibnamefont{Sullivan}},
  \bibinfo{author}{\bibfnamefont{W.}~\bibnamefont{Pan}},
  \bibinfo{author}{\bibfnamefont{K.~W.} \bibnamefont{Baldwin}},
  \bibinfo{author}{\bibfnamefont{K.~W.} \bibnamefont{West}},
  \bibinfo{author}{\bibfnamefont{L.~N.} \bibnamefont{Pfeiffer}},
  \bibnamefont{and} \bibinfo{author}{\bibfnamefont{D.~C.} \bibnamefont{Tsui}},
  \bibinfo{journal}{Phys. Rev. B} \textbf{\bibinfo{volume}{85}},
  \bibinfo{pages}{241302(R)} (\bibinfo{year}{2012}),
  \urlprefix\url{https://link.aps.org/doi/10.1103/PhysRevB.85.241302}.

\bibitem[{\citenamefont{Deng et~al.}(2012)\citenamefont{Deng, Watson,
  Rokhinson, Manfra, and Cs\'athy}}]{Deng12}
\bibinfo{author}{\bibfnamefont{N.}~\bibnamefont{Deng}},
  \bibinfo{author}{\bibfnamefont{J.~D.} \bibnamefont{Watson}},
  \bibinfo{author}{\bibfnamefont{L.~P.} \bibnamefont{Rokhinson}},
  \bibinfo{author}{\bibfnamefont{M.~J.} \bibnamefont{Manfra}},
  \bibnamefont{and} \bibinfo{author}{\bibfnamefont{G.~A.}
  \bibnamefont{Cs\'athy}}, \bibinfo{journal}{Phys. Rev. B}
  \textbf{\bibinfo{volume}{86}}, \bibinfo{pages}{201301(R)}
  (\bibinfo{year}{2012}),
  \urlprefix\url{https://link.aps.org/doi/10.1103/PhysRevB.86.201301}.

\bibitem[{\citenamefont{Balram and Jain}(2017)}]{Balram17}
\bibinfo{author}{\bibfnamefont{A.~C.} \bibnamefont{Balram}} \bibnamefont{and}
  \bibinfo{author}{\bibfnamefont{J.~K.} \bibnamefont{Jain}},
  \bibinfo{journal}{Phys. Rev. B} \textbf{\bibinfo{volume}{96}},
  \bibinfo{pages}{245142} (\bibinfo{year}{2017}),
  \urlprefix\url{https://link.aps.org/doi/10.1103/PhysRevB.96.245142}.

\bibitem[{\citenamefont{Sreejith et~al.}(2017)\citenamefont{Sreejith, Zhang,
  and Jain}}]{Sreejith2017}
\bibinfo{author}{\bibfnamefont{G.~J.} \bibnamefont{Sreejith}},
  \bibinfo{author}{\bibfnamefont{Y.}~\bibnamefont{Zhang}}, \bibnamefont{and}
  \bibinfo{author}{\bibfnamefont{J.~K.} \bibnamefont{Jain}},
  \bibinfo{journal}{Phys. Rev. B} \textbf{\bibinfo{volume}{96}},
  \bibinfo{pages}{125149} (\bibinfo{year}{2017}),
  \urlprefix\url{https://link.aps.org/doi/10.1103/PhysRevB.96.125149}.

\bibitem[{\citenamefont{Pan et~al.}(2020)\citenamefont{Pan, Kang, Lilly, Reno,
  Baldwin, West, Pfeiffer, and Tsui}}]{Pan20}
\bibinfo{author}{\bibfnamefont{W.}~\bibnamefont{Pan}},
  \bibinfo{author}{\bibfnamefont{W.}~\bibnamefont{Kang}},
  \bibinfo{author}{\bibfnamefont{M.~P.} \bibnamefont{Lilly}},
  \bibinfo{author}{\bibfnamefont{J.~L.} \bibnamefont{Reno}},
  \bibinfo{author}{\bibfnamefont{K.~W.} \bibnamefont{Baldwin}},
  \bibinfo{author}{\bibfnamefont{K.~W.} \bibnamefont{West}},
  \bibinfo{author}{\bibfnamefont{L.~N.} \bibnamefont{Pfeiffer}},
  \bibnamefont{and} \bibinfo{author}{\bibfnamefont{D.~C.} \bibnamefont{Tsui}},
  \bibinfo{journal}{Phys. Rev. Lett.} \textbf{\bibinfo{volume}{124}},
  \bibinfo{pages}{156801} (\bibinfo{year}{2020}),
  \urlprefix\url{https://link.aps.org/doi/10.1103/PhysRevLett.124.156801}.

\bibitem[{\citenamefont{Haldane}(1983)}]{Haldane83}
\bibinfo{author}{\bibfnamefont{F.~D.~M.} \bibnamefont{Haldane}},
  \bibinfo{journal}{Phys. Rev. Lett.} \textbf{\bibinfo{volume}{51}},
  \bibinfo{pages}{605} (\bibinfo{year}{1983}),
  \urlprefix\url{https://link.aps.org/doi/10.1103/PhysRevLett.51.605}.

\bibitem[{\citenamefont{Jain}(1989)}]{Jain89}
\bibinfo{author}{\bibfnamefont{J.~K.} \bibnamefont{Jain}},
  \bibinfo{journal}{Phys. Rev. Lett.} \textbf{\bibinfo{volume}{63}},
  \bibinfo{pages}{199} (\bibinfo{year}{1989}),
  \urlprefix\url{https://link.aps.org/doi/10.1103/PhysRevLett.63.199}.

\bibitem[{\citenamefont{Wen}(1990)}]{Wen90a}
\bibinfo{author}{\bibfnamefont{X.~G.} \bibnamefont{Wen}},
  \bibinfo{journal}{Int. J. Mod. Phys. B} \textbf{\bibinfo{volume}{4}},
  \bibinfo{pages}{239} (\bibinfo{year}{1990}),
  \urlprefix\url{https://doi.org/10.1142/S0217979290000139}.

\bibitem[{\citenamefont{Halperin et~al.}(1993)\citenamefont{Halperin, Lee, and
  Read}}]{Halperin93}
\bibinfo{author}{\bibfnamefont{B.~I.} \bibnamefont{Halperin}},
  \bibinfo{author}{\bibfnamefont{P.~A.} \bibnamefont{Lee}}, \bibnamefont{and}
  \bibinfo{author}{\bibfnamefont{N.}~\bibnamefont{Read}},
  \bibinfo{journal}{Phys. Rev. B} \textbf{\bibinfo{volume}{47}},
  \bibinfo{pages}{7312} (\bibinfo{year}{1993}),
  \urlprefix\url{https://link.aps.org/doi/10.1103/PhysRevB.47.7312}.

\bibitem[{\citenamefont{Morf and d'Ambrumenil}(1995)}]{Morf1995PRL}
\bibinfo{author}{\bibfnamefont{R.}~\bibnamefont{Morf}} \bibnamefont{and}
  \bibinfo{author}{\bibfnamefont{N.}~\bibnamefont{d'Ambrumenil}},
  \bibinfo{journal}{Phys. Rev. Lett.} \textbf{\bibinfo{volume}{74}},
  \bibinfo{pages}{5116} (\bibinfo{year}{1995}),
  \urlprefix\url{https://link.aps.org/doi/10.1103/PhysRevLett.74.5116}.

\bibitem[{\citenamefont{Morf et~al.}(2002)\citenamefont{Morf, d'Ambrumenil, and
  Das~Sarma}}]{Morf02}
\bibinfo{author}{\bibfnamefont{R.~H.} \bibnamefont{Morf}},
  \bibinfo{author}{\bibfnamefont{N.}~\bibnamefont{d'Ambrumenil}},
  \bibnamefont{and}
  \bibinfo{author}{\bibfnamefont{S.}~\bibnamefont{Das~Sarma}},
  \bibinfo{journal}{Phys. Rev. B} \textbf{\bibinfo{volume}{66}},
  \bibinfo{pages}{075408} (\bibinfo{year}{2002}),
  \urlprefix\url{https://link.aps.org/doi/10.1103/PhysRevB.66.075408}.

\bibitem[{\citenamefont{Fang and Howard}(1966)}]{FangPRL1966}
\bibinfo{author}{\bibfnamefont{F.~F.} \bibnamefont{Fang}} \bibnamefont{and}
  \bibinfo{author}{\bibfnamefont{W.~E.} \bibnamefont{Howard}},
  \bibinfo{journal}{Phys. Rev. Lett.} \textbf{\bibinfo{volume}{16}},
  \bibinfo{pages}{797} (\bibinfo{year}{1966}),
  \urlprefix\url{https://link.aps.org/doi/10.1103/PhysRevLett.16.797}.

\bibitem[{\citenamefont{Ando et~al.}(1982)\citenamefont{Ando, Fowler, and
  Stern}}]{AndoRMP1982}
\bibinfo{author}{\bibfnamefont{T.}~\bibnamefont{Ando}},
  \bibinfo{author}{\bibfnamefont{A.~B.} \bibnamefont{Fowler}},
  \bibnamefont{and} \bibinfo{author}{\bibfnamefont{F.}~\bibnamefont{Stern}},
  \bibinfo{journal}{Rev. Mod. Phys.} \textbf{\bibinfo{volume}{54}},
  \bibinfo{pages}{437} (\bibinfo{year}{1982}),
  \urlprefix\url{https://link.aps.org/doi/10.1103/RevModPhys.54.437}.

\bibitem[{\citenamefont{Stern and Das~Sarma}(1984)}]{SternPRB1984}
\bibinfo{author}{\bibfnamefont{F.}~\bibnamefont{Stern}} \bibnamefont{and}
  \bibinfo{author}{\bibfnamefont{S.}~\bibnamefont{Das~Sarma}},
  \bibinfo{journal}{Phys. Rev. B} \textbf{\bibinfo{volume}{30}},
  \bibinfo{pages}{840} (\bibinfo{year}{1984}),
  \urlprefix\url{https://link.aps.org/doi/10.1103/PhysRevB.30.840}.

\bibitem[{\citenamefont{Peterson et~al.}(2008)\citenamefont{Peterson,
  Jolicoeur, and Das~Sarma}}]{Peterson08b}
\bibinfo{author}{\bibfnamefont{M.~R.} \bibnamefont{Peterson}},
  \bibinfo{author}{\bibfnamefont{T.}~\bibnamefont{Jolicoeur}},
  \bibnamefont{and}
  \bibinfo{author}{\bibfnamefont{S.}~\bibnamefont{Das~Sarma}},
  \bibinfo{journal}{Phys. Rev. B} \textbf{\bibinfo{volume}{78}},
  \bibinfo{pages}{155308} (\bibinfo{year}{2008}),
  \urlprefix\url{https://link.aps.org/doi/10.1103/PhysRevB.78.155308}.

\bibitem[{\citenamefont{Park and Jain}(1998)}]{Park98}
\bibinfo{author}{\bibfnamefont{K.}~\bibnamefont{Park}} \bibnamefont{and}
  \bibinfo{author}{\bibfnamefont{J.~K.} \bibnamefont{Jain}},
  \bibinfo{journal}{Phys. Rev. Lett.} \textbf{\bibinfo{volume}{81}},
  \bibinfo{pages}{4200} (\bibinfo{year}{1998}),
  \urlprefix\url{https://link.aps.org/doi/10.1103/PhysRevLett.81.4200}.

\bibitem[{\citenamefont{Zhang et~al.}(2016)\citenamefont{Zhang, W\'ojs, and
  Jain}}]{Yuhe2016}
\bibinfo{author}{\bibfnamefont{Y.}~\bibnamefont{Zhang}},
  \bibinfo{author}{\bibfnamefont{A.}~\bibnamefont{W\'ojs}}, \bibnamefont{and}
  \bibinfo{author}{\bibfnamefont{J.~K.} \bibnamefont{Jain}},
  \bibinfo{journal}{Phys. Rev. Lett.} \textbf{\bibinfo{volume}{117}},
  \bibinfo{pages}{116803} (\bibinfo{year}{2016}),
  \urlprefix\url{https://link.aps.org/doi/10.1103/PhysRevLett.117.116803}.

\bibitem[{\citenamefont{Balram and Jain}(2016)}]{Balram16}
\bibinfo{author}{\bibfnamefont{A.~C.} \bibnamefont{Balram}} \bibnamefont{and}
  \bibinfo{author}{\bibfnamefont{J.~K.} \bibnamefont{Jain}},
  \bibinfo{journal}{Phys. Rev. B} \textbf{\bibinfo{volume}{93}},
  \bibinfo{pages}{235152} (\bibinfo{year}{2016}),
  \urlprefix\url{https://link.aps.org/doi/10.1103/PhysRevB.93.235152}.

\end{thebibliography}
\end{document}